\documentclass{LMCS}

\usepackage{ifthen}
\newboolean{full}
\setboolean{full}{true}

\usepackage[all]{xy}
\SelectTips{cm}{}
\usepackage{amsbsy,amssymb,eucal,rotating,style,array,longtable,hyperref}

\theoremstyle{definition}

\newtheorem{construction}[thm]{Construction}
\newtheorem{notation}[thm]{Notation}

\newenvironment{remarkNoNumber}{%
  \par\noindent
  {\bf{Remark.}}\rm~}
{\par}

\def\motto{%
  \vspace{1ex}
  \begin{flushright}
    \begin{minipage}{8cm}
      \begin{flushright}
        \footnotesize
        If you are not part of the solution, \\
        you are part of the problem.
        
        \smallskip
        Eldridge Cleaver, {\em speech in San Francisco}, 1968
      \end{flushright}
    \end{minipage}
  \end{flushright}
  \vspace{1ex}
  }

\newcommand{\comment}[1]{\empty}

\def\doi{2 (5:4) 2006}
\lmcsheading%
{\doi}
{1--31}
{}
{}
{Mar.~10, 2006}
{Nov.~\phantom{0}8, 2006}
{} 
  
\begin{document}

%
%

\title{Elgot Algebras\rsuper{\dagger}}
\titlecomment{{\lsuper{\dagger}}This paper is a full version of an
extended abstract~\cite{abs} presented at the conference MFPS XXI.}

\author[J.~Ad\'{a}mek]{Ji\v{r}\'{\i} Ad\'{a}mek\rsuper{a}}
\address{{\lsuper{a,b}}Institute of Theoretical Computer Science, 
  Technical University, Braunschweig, Germany}
\email{\{adamek,milius\}@iti.cs.tu-bs.de}
\thanks{{\lsuper{a,c}}The first and the third author acknowledge the
  support of the Grant Agency of the Czech Republic under the Grant
  No.~201/02/0148.}
\author[S.~Milius]{Stefan Milius\rsuper{b}}
\address{\vskip-6 pt}
%
\author[J.~Velebil]{Ji\v{r}\'{\i} Velebil\rsuper{c}}
\address{{\lsuper{c}}Faculty of Electrical Engineering,
         Technical University, Prague, Czech Republic}
\email{velebil@math.feld.cvut.cz}

\keywords{Elgot algebra, rational monad, coalgebra, iterative theories}
\subjclass{F.3.2}

%
%

\begin{abstract}
  Denotational semantics can be based on algebras with additional structure
  (order, metric, etc.) which makes it possible to interpret recursive
  specifications. It was the idea of Elgot
  to base denotational semantics on iterative theories instead,
  i.\,e., theories in
  which abstract recursive specifications are required to have unique
  solutions. Later Bloom and \'Esik studied iteration theories and iteration
  algebras in which a specified solution has to obey certain axioms.
  We propose so-called Elgot algebras as a convenient structure
  for semantics in the present paper. An Elgot algebra is an
  algebra with a specified solution for every system of flat recursive
  equations. That specification satisfies two simple and well motivated axioms:
  functoriality (stating that solutions are stable under renaming of recursion
  variables) and compositionality (stating how to perform simultaneous
  recursion).  These two axioms stem canonically from Elgot's iterative
  theories: We prove that the category of Elgot algebras is the
  Eilenberg--Moore category of the monad given by a free iterative theory.
\end{abstract}

\maketitle

\motto
\section{Introduction}
\label{sec:1}
\setcounter{equation}{0}

We study Elgot algebras, a new notion of algebra
useful for application in the semantics of recursive computations. 
In programming, functions are often specified by a 
\emph{recursive program scheme} such as
\begin{equation}\label{eq:rps}
  \begin{array}{rcl}
    \phi(x) & \approx & F(x, \phi(Gx)) \\
    \psi(x) & \approx & F(\phi(Gx), GGx)
  \end{array}
\end{equation}
where $F$ and $G$ are given functions and $\phi$ and $\psi$ are recursively
defined in terms of the given ones by~\refeq{eq:rps}. We are interested in the
semantics of such schemes. Actually, one has to distinguish between
\emph{uninterpreted} and \emph{interpreted} semantics. In the uninterpreted
semantics the givens are not functions but merely function symbols from a
signature $\Sigma$. In the present paper we prepare a basis for the
interpreted semantics in which a program scheme comes together with a suitable
$\Sigma$-algebra $A$, which gives an interpretation to all the given function
symbols. The actual application of Elgot algebras to semantics will be dealt
with in joint work of the second author with Larry Moss~\cite{mm}.
By ``suitable algebra'' we mean, of course, one in which recursive
program schemes can be given a semantics. For example, for the recursive
program scheme~\refeq{eq:rps} we are only interested in those $\Sigma$-algebras
$A$, where $\Sigma = \{\,F,G\,\}$, in which the program scheme~\refeq{eq:rps}
has a \emph{solution}, i.\,e., we can canonically obtain new operations
$\phi^A$ and $\psi^A$ on $A$ so that the formal equations~\refeq{eq:rps} become
valid identities.  The question we address is:
\begin{equation}\label{eq:prob}
  \textrm{
    What $\Sigma$-algebras are suitable for semantics?
    }
\end{equation}
Several answers have been proposed in the literature. One well-known approach
is to work with complete posets (CPO) in lieu of sets, see e.g.~\cite{adj}.
Here algebras have an additional CPO structure making all operations
continuous. Another approach works with complete metric spaces, see
e.g.~\cite{aru}. Here we have an additional complete metric making all
operations contracting. In both of these approaches one imposes extra structure
on the algebra in a way that makes it possible to obtain the semantics of a
recursive computation as a join (or limit, respectively) of finite
approximations.

It was the idea of Calvin Elgot to try and work in a purely algebraic setting
avoiding extra structure like order or metric. In~\cite{e} he introduced
iterative theories which are algebraic theories in which certain systems of
recursive equations have \emph{unique} solutions. Later Evelyn Nelson~\cite{n}
and Jerzy Tiuryn~\cite{t} studied iterative algebras, which are algebras for a
signature $\Sigma$ with unique solutions of recursive equations. While avoiding
extra structure, these are still not the unifying concept one would hope for,
since they do not subsume continuous algebras---least fixed points are
typically not unique.

However, analyzing all the above types of algebras we find an interesting
common feature which makes continuous, metrizable and iterative algebras fit for
use in semantics of recursive program schemes: these algebras allow for an
interpretation of infinite $\Sigma$-trees. Let us make this more precise. For a
given signature $\Sigma$ consider the algebra
$$
T_\Sigma X
$$
of all (finite and infinite) $\Sigma$-trees over $X$, i.\,e., rooted ordered
trees where inner nodes with $n$ children are labelled by $n$-ary operation
symbols from $\Sigma$, and leaves are labelled by constants or elements from
$X$. It is well-known that for any continuous (or metrizable)
algebra $A$ there is a unique continuous (or contracting, respectively)
homomorphism from $T_\Sigma A$ to $A$ which provides for any $\Sigma$-tree over $A$
its result of computation in $A$. It is then easy to give semantics to
recursive program schemes in $A$. For example, for~\refeq{eq:rps} one can
simply take the tree unfolding which yields the infinite trees
$$
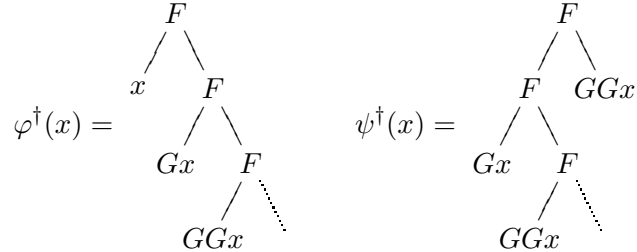
\sol{\phi}(x) = 
\vcenter{
  \xy
  \POS (00,00) *+{F} = "A"
     , (-5,-10) *+{x} = "x"
     , (5,-10) *+{F} = "A1"
     , (0,-20) *+{Gx} = "Bx"
     , (10,-20) *+{F} = "A2"
     , (5,-30)  *+{GGx} = "BBx"
     , (15,-30) *+{} = "end"
  \ar@{-} "A";"x"
  \ar@{-} "A";"A1"
  \ar@{-} "A1";"Bx" 
  \ar@{-} "A1";"A2"
  \ar@{-} "A2";"BBx"
  \ar@{.} "A2";"end"
  \endxy
}
\qquad
\sol{\psi}(x) =
\vcenter{
  \xy
  \POS (000,000)  *+{F}   = "A"
     , ( -5,-10)  *+{F}   = "F"
     , (  5,-10)  *+{GGx} = "BBx"
     , (-10,-20)  *+{Gx}  = "Gx"
     , (  0,-20)  *+{F}   = "F2"
     , ( -5,-30)  *+{GGx} = "GGx"
     , (  5,-30)  *+{}    = "end"
  \ar@{-} "A";"F"
  \ar@{-} "A";"BBx"
  \ar@{-} "F";"Gx"
  \ar@{-} "F";"F2"
  \ar@{-} "F2";"GGx"
  \ar@{.} "F2";"end"
  \endxy
}
$$
and then for any argument $x \in A$ compute these infinite trees in $A$. 

Actually, we do not need to be able to compute all infinite trees: all
recursive program schemes unfold to \emph{algebraic trees}, see~\cite{c}
(we discuss these briefly in Section~\ref{sec:6} below).
Another important subclass are
\emph{rational trees}, which are obtained as all solutions of guarded finitary
recursive equations. They were characterized
by Sussanna Ginali~\cite{g} as those
$\Sigma$-trees having up to isomorphism finitely many subtrees only. We denote
by
$$
R_\Sigma X
$$
the subalgebra of all rational trees in $T_\Sigma X$. With this in mind, we can
restate problem~\refeq{eq:prob} more formally:
\begin{equation}\label{eq:prob2}
  \textrm{%
    \parbox{11cm}{
      \hfill What $\Sigma$-algebras have a suitable computation of all trees? 
      \hfill \\
      \hspace*{\fill} Or all rational trees? \hspace*{\fill}
      }
    }
\end{equation}
This means, one further step more formally: what is the largest category of
$\Sigma$-algebras in which $T_\Sigma X$, or $R_\Sigma X$, respectively, act as
free algebras on $X$? The answer in the case of $T_\Sigma X$ is: complete Elgot
algebras. These are $\Sigma$-algebras $A$ with an additional operation 
``dagger'' assigning to every system $e$ of recursive equations in $A$ a
solution $\sol{e}$. Two (surprisingly simple) axioms are put on $\solop$ which
stem from the internal structure of $T_\Sigma X$: the functor $T_\Sigma$ given
by $X \mapsto T_\Sigma X$ is part of a monad on $\Set$, and this
monad yields the free
completely iterative theory on $\Sigma$, as proved in~\cite{ebt}. We will prove
that the algebras for this monad (i.\,e., the Eilenberg--Moore category
of $T_\Sigma$) are complete Elgot algebras.  Basic
examples: continuous algebras or metrizable algebras are
complete Elgot algebras.
Analogously, the largest category of $\Sigma$-algebras in which each $R_\Sigma
X$ acts as a free algebra
is the category of Elgot algebras. They are defined precisely as the
complete Elgot algebras, except that the systems $e$ of recursive equations
considered there are required to be finite. For example, every iterative
algebra is an Elgot algebra. We present the results for
suitable endofunctors of an
arbitrary category $\A$ satisfying very mild conditions: for complete Elgot
algebras we just need $\A$ to have finite coproducts, for Elgot algebras we
work with locally finitely presentable categories in the sense of
Peter Gabriel and Friedrich Ulmer~\cite{ar}.

\medskip\noindent
{\bf Related Work:} 
Solutions of recursive equations are a fundamental part of a number of models
of computation, e.\,g., iterative theories of C.~Elgot~\cite{e}, iteration
theories of S.~Bloom and Z.~\'Esik~\cite{be}, traced monoidal categories of
A.~Joyal, R.~Street and D.~Verity~\cite{jsv}, fixed-point theories for domains,
see S.~Eilenberg~\cite{ei} or G.~Plotkin~\cite{p}, etc. In some of these models
the assignment of a solution $\sol{e}$ to a given type of recursive equation
$e$ is unique (e.\,g., in iterative theories every ideal system has a unique
solution, or in domains given by a complete metric space there are unique
solutions of fixed-point equations, see~\cite{aru}). The operation $e \mapsto
\sol{e}$ then satisfies a number of equational properties. In other models,
like in iteration theories, for example,
a specific choice of a solution $\sol{e}$ is
assumed, and certain properties (inspired by the models with unique solutions)
are formulated as axioms. Recall that in a traced monoidal category whose
tensor product is just the ordinary product the trace is equivalently presented
in form of an operation $\sol{e}$ satisfying certain axioms, see~\cite{h}
and~\cite{hagh}. 

The approach of the present paper is more elementary in asking for solutions $e
\mapsto \sol{e}$ in a concrete algebra $A$. Here we work with flat equations
$e$ in $A$, which are morphisms of the form $e: X \to HX + A$.
However, flatness is
just a technical restriction: in future research we will prove that more
general non-flat equations obtain solutions ``automatically''. The fact that we
work with a fixed algebra $A$ (and let only $X$ and $e$ vary) is partly
responsible for the simplicity of our axioms in comparison to the work on
theories (where $A$ varies as well), see e.\,g.~\cite{be} or~\cite{simp}.
Iterative algebras of Evelyn Nelson~\cite{n} and Jerzy Tiuryn~\cite{t}, where
solutions $\sol{e}$ are required to be unique, are a similar approach.
Furthermore, iteration algebras of Zoltan \'{E}sik~\cite{esik} are another one.
Unfortunately, the number of axioms (seven) and their complexity make the
question of the relationship of that notion to Elgot algebras a nontrivial one.
We intend to study this question in the future.

\sloppypar
We work with two variations: Elgot algebras, related to $R_\Sigma X$, where the
function $\solop$ assigns a solution only to finitary flat recursive systems,
and complete Elgot algebras, related to $T_\Sigma X$, where the function
$\solop$ assigns solutions to all flat recursive systems. This is based on
our previous research~\cite{aamv,m,amv1,amv} in which we proved that every finitary
endofunctor $H$ generates a free iterative monad $R$, and a free completely
iterative monad $T$. In the present paper we study the Eilenberg--Moore
categories of the monads $R$ and $T$.

\medskip\noindent
{\bf Organization of the Paper:}
In Section~\ref{sec:2} we recall (completely) iterative
algebras and prove that the assignment of unique solutions
in these algebras fulfills the axioms of functoriality and
compositionality.

Elgot algebras and complete Elgot algebras  are introduced
in Section~\ref{sec:3}
as algebras equipped with a chosen assignment of
a solution that satisfies functoriality and compositionality.

In Sections~\ref{sec:4} and~\ref{sec:5} we prove that
(complete) Elgot algebras form Eilenberg--Moore category
of a free (completely) iterative monad.
\iffull\else
We omit some proofs for lack of space, the reader can find them in the 
full version of this paper~\cite{full}. 
\fi

\section{Iterative Algebras and CIAs}
\label{sec:2}
\setcounter{equation}{0}

\begin{asm}
  Throughout the paper $H$ denotes an endo\-func\-tor of a category $\kat{A}$
  having binary coproducts. We denote the corresponding injections
  by $\inl: A \to A+B$ and $\inr: B \to A+B$.
\end{asm}

Recall that an object $X$ of a category with filtered colimits
is called \emph{finitely presentable} if the
$\hom$-functor $\kat{A}(X,-): \kat{A} \to \Set$ is finitary,
i.\ e., if it preserves filtered colimits. (In $\Set$, these
are precisely the finite sets. In equational classes of algebras these are
precisely the finitely presentable algebras in the usual sense.) Recall further
that a category $\A$ is called \emph{locally finitely presentable} if it has
colimits and a set of finitely presentable objects whose
closure under filtered colimits is all of $\A$, see~\cite{ar}. (Examples:
the categories of sets, posets, graphs or any finitary variety of algebras are
locally finitely presentable categories.)

\begin{defi}
\label{def:2.1}
  By a \emph{flat equation morphism}
  in an object $A$ we understand a morphism $e: X \to HX + A$
  in $\kat{A}$. We call $e$
  \emph{finitary} provided that $X$ is finitely
  presentable.\footnote{We shall only use this notion in the case
  when $\kat{A}$ is locally finitely presentable and $H$ is finitary.}
  Suppose that $A$ is the carrier of an $H$-algebra
  $\alpha:HA\to A$.
  A \emph{solution}
  of $e$ is a morphism $\sol{e}: X \to A$ such that the square
  \begin{equation}\label{diag:sol}
    \vcenter{
      \xymatrix{
        X \ar[d]_e \ar[rr]^-{\sol{e}} &  & A \\
        HX + A \ar[rr]_-{H\sol{e} + A} & & HA + A \ar[u]_{[\alpha, A]}
      }
  }
  \end{equation}
  commutes.
  
  If every finitary flat equation morphism has a unique solution, then $A$ is
  said to be an \emph{iterative algebra}. The algebra $A$ is
  called a \emph{completely
  iterative algebra} (CIA) if every flat equation morphism has a unique
  solution.
\end{defi}

\begin{rem}
  Iterative algebras for polynomial endofunctors of $\Set$ were introduced and
  studied by Evelyn Nelson~\cite{n}. She proved that the algebras $R_\Sigma X$
  of rational $\Sigma$-trees on $X$ are free iterative algebras, and that the
  algebraic theory obtained from them is a free iterative theory of Calvin
  Elgot~\cite{e}. We have recently studied iterative algebras in a much more
  general setting, working with a finitary endofunctor of a locally finitely
  presentable category. Completely iterative algebras were studied by Stefan
  Milius~\cite{m}.
\end{rem}

\begin{exa}
  Consider algebras in $\Set$ with one binary operation $*$.
  In that case, the
  functor is $HX = X \times X$. A flat equation morphism $e$ in an algebra $A$
  assigns to every variable $x$ either a flat term $y * z$ ($y$ and $z$ are
  variables) or an element of $A$. A solution $\sol{e}: X \to A$ assigns to $x
  \in X$ either the same element as $e$, in case $e(x) \in A$, or the result
  of $\sol{e}(y) * \sol{e}(z)$, in case $e(x) = y * z$. For example, the
  following recursive equation
  $$
  x \approx x * x\,,
  $$
  represented by the obvious morphism $e: \{\,x\,\} \to \{\,x\,\} \times
  \{\,x\,\} + A$, has as solution $\sol{e}$ an element $a = \sol{e}(x)$ which
  is idempotent. Consequently, every iterative algebra has a unique idempotent.
  If $A$ is even completely iterative, then it has, for each sequence
  $a_0, a_1, a_2, \ldots$ of elements, a unique interpretation of $a_0 * (a_1
  * (a_2\cdots)))$, i.\,e., a unique sequence $b_0, b_1, b_2, \ldots$ with
  $b_0 = a_0 * b_1$, $b_1 = a_1 * b_2$, etc. In fact, we consider here the
  equations 
  $$
  x_n \approx a_n * x_{n+1} \qquad (n \in \Nat)\,.
  $$

  Iterative algebras have unique solutions of many non-flat equations because
  we can flatten them. For example the following recursive equations
  $$
  \begin{array}{rcl@{\qquad}rcl}
    x_1 & \approx & (x_2 * a) * b &
    x_2 & \approx & x_1 * b
  \end{array}
  $$
  are not flat. But they can be easily flattened to obtain a system
  $$
  \begin{array}{rcl@{\qquad}rcl}
    x_1 & \approx & z_1 * z_2 & x_2 & \approx & x_1 * z_2 \\
    z_1 & \approx & x_2 * z_3 & z_2 & \approx & b  \\
    z_3 & \approx & a \\
  \end{array}
  $$
  represented by a morphism $e: X \to X \times X + A$, where $X = \{\,x_1, x_2,
  z_1, z_2, z_3\,\}$. Its solution is a map $\sol{e}: X \to A$ yielding a pair
  of elements $s = \sol{e}(x_1)$ and $t = \sol{e}(x_2)$ satisfying $s = (t *
  a) * b$ and $t = s * b$.
\end{exa}

\begin{exa}
{\em Iterative $\Sigma$-algebras.}
For every finitary signature $\Sigma=(\Sigma_n)_{n\in\Nat}$
we can identify $\Sigma$-algebras with algebras for the
{\em polynomial endofunctor} $H_\Sigma$ of $\Set$ defined on objects $X$ by 
$$
H_\Sigma X = \Sigma_0+\Sigma_1\times X+\Sigma_2\times X\times X+\dots
$$
A $\Sigma$-term which has the form
$\sigma(x_{1},\dots,x_{k})$ for some $\sigma\in\Sigma_k$ and for variables
$x_1, \ldots, x_k$ from $X$ is
called {\em flat}. Then a flat equation morphism $e:X\to H_\Sigma X+A$
in an algebra $A$ represents a system
$$
x \approx t_x
$$
of recursive equations, one for every variable $x \in X$, where each $t_x$ is
either a flat term in $X$, or an element of $A$. A solution $\sol{e}$
assigns to every variable $x$ with $t_x = a$, $a \in A$, the element $a$, and
if $t_x = \sigma(x_1, \ldots, x_k)$ then $\sol{e}(x) =
\sigma_A(\sol{e}(x_1),\ldots, \sol{e}(x_k))$. 

Observe that every iterative $\Sigma$-algebra $A$ has, for every
$\sigma\in\Sigma_k$, a unique idempotent (i.e., a unique element
$a\in A$ with $\sigma(a,\dots,a)=a$). In fact, consider the flat
equation $x\approx\sigma(x,\dots,x)$. More generally, every
$\Sigma$-term has a unique idempotent in $A$. For example,
for a term of depth $2$, $\sigma(\tau_1, \ldots, \tau_k)$,
where $\sigma\in\Sigma_k$ and $\tau_1,\dots,\tau_k\in\Sigma_n$
consider the recursive equations
\begin{eqnarray*}
x_0 &\approx& \sigma(x_1,x_2,\dots,x_k) \\
x_i &\approx& \tau_i(x_0,x_0,\dots,x_0) \qquad (i = 1, \ldots, k)\,. \\
\end{eqnarray*}
An example of an iterative $\Sigma$-algebra
is the algebra $T_\Sigma$ of all (finite and infinite) 
$\Sigma$-trees. Also the subalgebra $R_\Sigma$ of $T_\Sigma$
of all rational $\Sigma$-trees is iterative, see~\cite{n}.
\end{exa}

\begin{exa}
\label{ex:2.2}
In particular, for unary algebras ($H=\Id$), an algebra $\alpha:A\to A$ is
iterative iff $\alpha^k$ has a unique fixed point ($k\geq 1$), see~\cite{amv}.  
The algebra $A$ is a CIA iff, in addition to a unique fixed point
of each $\alpha^k$, there exists no infinite sequence
$(a_n)_{n\in \Nat}$ in $\kat{A}$ with $\alpha a_{n+1} = a_n$, see~\cite{m}.
\end{exa}
\begin{rem}
  \label{rem:rat}
  In~\cite{amv} we have proved that for every finitary functor $H$ of a locally
  finitely presentable category $\A$, a free iterative algebra $RY$ exists on
  every object $Y$.  Furthermore, we have given a canonical construction of
  $RY$ as a colimit of all coalgebras $X \to HX + Y$ carried by finitely
  presentable objects, in other words, for every object $Y$ of $\A$, $RY$ is a
  colimit of all finitary flat equations in $Y$.  For example, for a polynomial
  functor $H_\Sigma$ of $\Set$ the free iterative algebra on a set $Y$ is the
  algebra $R_\Sigma Y$ of all rational $\Sigma$-trees over $Y$.  In general, we
  call the monad $\R$ of free iterative algebras the \emph{rational monad}
  generated by $H$.  We have proved in~\cite{amv} that the rational monad $\R$
  is a free iterative monad on $H$.
\end{rem}

\begin{exa}
\label{ex:cms}
{\it Completely metrizable algebras.} 
Complete metric spaces are well-known to be a suitable basis for semantics. The
first categorical treatment of complete metric spaces for semantics is due to 
Pierre America and Jan Rutten~\cite{aru}. Let
$$
\CMS
$$
denote the category of all complete metric spaces (i.e., such that every
Cauchy sequence has a limit) with metrics in the interval $[0,1]$. The
morphisms are maps $f:(X,d_X)\to (Y,d_Y)$ where the inequality
$d_Y(f(x),f(x'))\leq d_X(x,x')$ holds for all $x$, $x'$ in $X$.
Such maps $f$ are called \emph{nonexpanding}.

Given complete metric spaces $X$ and $Y$, the hom-set $\CMS(X,Y)$ carries the
pointwise metric $d_{X,Y}$ defined as follows:
$$
d_{X,Y}(f,g)=\sup_{x\in X} d_Y(f(x),g(x))
$$
America and Rutten call a functor $H:\CMS\to\CMS$ {\em contracting} if there
exists a constant $\eps<1$ such that for arbitrary morphisms
$f,g:X\to Y$ we have
\begin{equation}
\label{eq:2.0}
d_{HX,HY}(Hf,Hg)\leq\eps\cdot d_{X,Y}(f,g).
\end{equation}
\end{exa}

\begin{lem}
\label{lem:met}
If $H:\CMS\to\CMS$ is a contracting functor, then every nonempty
$H$-algebra is a CIA. 
\end{lem}
\begin{proof}
  Let $\alpha:HA\to A$ be a nonempty $H$-algebra.
  Recall that the hom-set $\CMS(X,A)$ is a complete
  metric space with the supremum metric. Definition~\ref{def:2.1}
  of a solution of a flat equation morphism $e:X\to HX+A$
  states that $\sol{e}$ is a fixed point of the function
  $F$ on $\CMS(X,A)$ given by
  $$
  F:(s:X\to A)\mapsto ([\alpha,A]\cdot (Hs+A)\cdot e).
  $$
  This function is a contraction on $\CMS(X,A)$. In fact,
  for two nonexpanding maps $s,t:X\to A$ we have
  $$
  \begin{array}{rclp{6.25cm}}
  d_{X,A}(Fs,Ft)
  &=&
  \multicolumn{2}{l}{%
    d_{X,A}([\alpha,A]\cdot (Hs+A)\cdot e,[\alpha,A]\cdot (Ht+A)\cdot e)
    }
  \\
  &&&
  \mbox{(by the definition of $F$)}
  \\
  &\leq&
  d_{HX,HA}(Hs+A,Ht+A)
  &
  \mbox{(since composition is nonexpanding)}
  \\
  &=&
  d_{HX,HA}(Hs,Ht)
  \\
  &\leq&
  \eps d_{X,A}(s,t)
  &
  \mbox{(since $H$ is contracting),}
  \end{array}
  $$
  where $\eps<1$ is the constant of~\refeq{eq:2.0} above.

  By Banach's Fixed Point Theorem, there exists a unique
  fixed point of $F$: a unique solution of $e$.
\end{proof}

\begin{rem}
\label{rem:metelgot}
\hfill
\begin{enumerate}
\item The proof of the last theorem yields a concrete formula
      for the unique solution $\sol{e}$ of a given flat
      equation morphism $e:X\to HX+A$. This unique solution
      is given as the limit of a Cauchy sequence in
      $\CMS(X,A)$ as follows:
      $$
      \sol{e}=\lim_{n\to\infty}\sol{e}_n,
      $$
      where $\sol{e}_0:X\to A$ is any nonexpanding map
      (for example a constant map: use that $A$ is nonempty)
      and $\sol{e}_{n+1}$ is defined by the commutativity of
      the diagram below:
      \begin{equation}
        \label{eq:approxsol}
        \vcenter{
          \xy
          \xymatrix{
            X
            \ar[0,2]^-{\sol{e}_{n+1}}
            \ar[1,0]_{e}
            &
            &
            A
            \\
            HX+A
            \ar[0,2]_-{H\sol{e}_n+A}
            &
            &
            HA+A
            \ar[-1,0]_{[\alpha,A]}
            }
          \endxy
          }
      \end{equation}
\item Many set functors $H$ have a lifting to contracting
      endofunctors $H'$ of $\CMS$. That is, for the forgetful functor
      $U:\CMS\to\Set$ the following square
      $$
      \xy
      \xymatrix{
      \CMS
      \ar[0,2]^-{H'}
      \ar[1,0]_{U}
      &
      &
      \CMS
      \ar[1,0]^{U}
      \\
      \Set
      \ar[0,2]_-{H}
      &
      &
      \Set
      }
      \endxy
      $$
      commutes. For example, if $HX= X^n$, define
      $$
      H'(X,d)=(X^n,\frac{1}{2}\cdot d'),
      $$
      where $d'$ is the maximum metric.
      Then $H'$ is a contracting functor with $\eps=\frac{1}{2}$.
      Since coproducts of $\frac{1}{2}$-contracting liftings are
      $\frac{1}{2}$-contracting liftings of coproducts, we conclude that
      every polynomial endofunctor has a contracting lifting to $\CMS$.
\end{enumerate}
\end{rem}

Let us call an $H$-algebra $\alpha:HA\to A$ {\em completely metrizable}
if there exists a complete metric, $d$, on $A$ such that
$\alpha$ is a nonexpanding map from $H'(A,d)$ to $(A,d)$.

\begin{cor}\label{cor:metrizable}
  Every completely metrizable algebra $A$ is a CIA.
\end{cor}
In fact, to every equation morphism $e:X\to HX+A$ assign the unique solution of
$e:(X,d_0)\to H'(X,d_0)+(A,d)$, where $d_0$ is the discrete metric ($d_0(x,x')=1$
iff $x\not= x'$).

\begin{rem}
  Stefan Milius~\cite{m} proved that for any endofunctor $H$ of $\A$ a final
  coalgebra $TY$ for $H(-)+Y$ is a free CIA on $Y$, and conversely. Furthermore,
  assuming that the free CIAs exist, it follows that the monad $\T$ of free
  CIAs is a free completely iterative monad on $H$. This generalizes and
  extends the classical result of
  Elgot, Bloom and Tindell~\cite{ebt} since for a polynomial functor
  $H_\Sigma$ of $\Set$ the free completely iterative algebra on a set $Y$ is
  the algebra $T_\Sigma Y$ of all $\Sigma$-trees over $Y$.
\end{rem}
\begin{rem}
\label{rem:2.7}
We are going to prove two properties of iterative algebras and CIA's:
functoriality and compositionality of solutions. We will use two
``operations'' on equation morphisms. One, $\after$, is just change of
parameter names: given a flat equation morphism $e:X\to HX+Y$ and a morphism
$h:Y\to Z$ we obtain the following equation morphism
$$
\xy
\xymatrix@C+1pc{
h\after e\equiv X
\ar[r]^-{e}
&
HX+Y
\ar[r]^-{HX+h}
&
HX+Z\,.
}
\endxy
$$
The other operation $\plus$ combines two flat equation morphisms
$$
e: X \to HX + Y
\qquad\textrm{and}\qquad
f: Y \to HY + A
$$
into the single flat equation morphism
$f \plus e: X + Y \to H(X+Y) + A$
in a canonical way: put $\can = [H\inl, H\inr]: HX + HY \to H(X+Y)$ and define 
\begin{equation}
  \label{eq:1.4}
  f\plus e\equiv 
  \let\objectstyle=\scriptstyle
  \let\labelstyle=\scriptscriptstyle
  \xymatrix@1{
    X+Y
    \ar[0,1]^-{[e,\inr]}
    &
    HX+Y
    \ar[0,1]^-{HX+f}
    &
    HX+HY+A
    \ar[0,1]^-{\can+A}
    &
    H(X+Y)+A,
    }
\end{equation}
\end{rem}

\subsection*{Functoriality}
  This states that solutions are invariant under renaming of variables,
  provided, of course, that the right-hand sides of equations are renamed
  accordingly. Formally, observe that every flat equation morphism is a
  coalgebra for the endofunctor $H({-})+A$. Given two such coalgebras $e$ and
  $f$, a renaming of the variables (or \emph{morphism of equations}) is a
  morphism $h: X \to Y$ which forms a coalgebra homomorphism:
\begin{equation}
  \label{eq:1.2}
  \vcenter{
    \xy
    \xymatrix{
      X
      \ar[0,2]^-{e}
      \ar[1,0]_{h}
      &
      &
      HX+A
      \ar[1,0]^{Hh+A}
      \\
      Y
      \ar[0,2]_-{f}
      &
      &
      HY+A
      }
    \endxy
    }
\end{equation}

\begin{defi}\label{dfn:func}
  Let $A$ be an algebra with a choice $e \mapsto \sol e$ of solutions, for all
  flat equation morphisms $e$ in $A$. We say that the choice is
  \emph{functorial} provided that 

  \begin{equation}\label{eq:func}
    \sol{e}=\sol{f}\cdot h
  \end{equation}
  holds for all morphisms $h: e \to f$ of equations.
  In other words: $\sol{({-})}$
  is a functor from the category of all flat equation morphisms in the algebra
  $A$ into the comma-category of the object $A$.
\end{defi}

\begin{lem}\label{lem:ciafunc}
  In every CIA the assignment $\solop$ is functorial. 
\end{lem}
\iffull
\begin{proof}
  For each morphism $h$ of equations the diagram 
  $$
  \xy
  \xymatrix{
    X
    \ar[0,2]_-{h}
    \ar[1,0]_{e}
    &
    &
    Y
    \ar[0,2]_-{\sol{f}}
    \ar[1,0]_{f}
    &
    &
    A
    \ar@{<-} `u[l] `[llll]_-{\sol{f}\cdot h} [llll]
    \\
    HX+A
    \ar[0,2]^-{Hh+A}
    &
    &
    HY+A
    \ar[0,2]^-{H\sol{f}+A}
    &
    &
    HA+A
    \ar[-1,0]_{[\alpha,A]}
    \ar@{<-} `d[l] `[llll]^-{H(\sol{f}\cdot h)+A} [llll]
  }
  \endxy
  $$
  commutes. Thus, $\sol f \cdot h$ is a solution of $e$. Uniqueness of
  solutions now implies the desired result.
\end{proof}
\fi
\begin{remarkNoNumber}
  The same holds for every iterative algebra, except that there we restrict
  $X$ and $Y$ in Definition~\ref{dfn:func} to finitely presentable objects. 
\end{remarkNoNumber}

\subsection*{Compositionality}
  This tells us how to perform simultaneous recursion: 
  given an equation morphism $f$ in $A$ with a variable object $Y$, we can
  combine it with any equation morphism $e$ in $Y$ with a variable object $X$
  to obtain the equation morphism $f \plus e$ in $A$ of Remark~\ref{rem:2.7}.
  Compositionality decrees that the left-hand component of
  $\sol{(f \plus e)}$ is just the solution of $\sol{f} \after e$.
  In other words: in
  lieu of solving $f$ and $e$ simultaneously we first solve $f$, plug in the
  solution in $e$ and solve the resulting equation morphism.

\begin{defi}\label{dfn:cia}
  Let $A$ be an algebra with a choice $e \mapsto \sol e$ of solutions, for all
  flat equation morphisms $e$ in $A$. We say that the choice is
  \emph{compositional} if for each pair $e: X \to HX + Y$ and $f: Y \to HY + A$
  of flat equation morphisms, we have

  \begin{equation}
    \label{eq:bs-identity}
    \sol{(\sol{f}\after e)}=
    \sol{(f\plus e)}\cdot\inl\,.
  \end{equation}
\end{defi}

\begin{rem}
  Notice that the coproduct injection $\inr: Y \to X + Y$ is a morphism of
  equations from $f$ to \mbox{$f \plus e$}.
  Functoriality then implies that $\sol{f} = \sol{(f \plus e)} \cdot \inr$. Thus,
  in the presence of functoriality, compositionality is equivalent to
  \begin{equation}\label{eq:modCIA}
    \sol{(f \plus e)} = [\sol{(\sol f \after e)}, \sol f]\,.
  \end{equation}
\end{rem}

\begin{lem}\label{lem:ciaid}
  In every CIA, the assignment $\solop$ is compositional.
\end{lem}
\iffull
\begin{proof}
Denote by
$$
r=\sol{(\sol{f}\after e)}:X\to A
$$
the solution of $\sol{f}\after e$. It is sufficient to prove that
the equation below holds:
$$
\sol{(f\plus e)}=[r,\sol{f}]:X+Y\to A\,.
$$
We establish this using the uniqueness of solutions and by showing
that the following diagram
\begin{equation}
\label{eq:2.1}
\vcenter{
\xy
\xymatrix{
& 
X+Y
\ar[0,2]^-{[r,\sol{f}]}
\ar[1,0]_{[e,\inr]}
\ar `l[dl] `[ddd]_{f \smallplus e} [ddd]
&
&
A
\\
&
HX+Y
\ar[1,0]_{HX+f}
&
&
\\
&
HX+HY+A
\ar[1,0]_{\can +A}
\ar[1,2]^*+{\labelstyle [Hr,H\sol{f}]+A}
&
&
\\
&
H(X+Y)+A
\ar[0,2]_-{H[r,\sol{f}]+A}
&
&
HA+A
\ar[-3,0]_{[\alpha,A]}
}
\endxy
}
\end{equation}
commutes. Commutation of the right-hand components (with domain $Y$)
of the diagram:
$$
[\alpha,A]\cdot ([Hr,H\sol{f}]+A)\cdot\inr\cdot f=
[\alpha,A]\cdot (H\sol{f}+A)\cdot f=
\sol{f}
$$
because $\sol{f}$ solves $f$. For the left-hand components (with domain $X$)
use the commutativity of the square defining $r=\sol{(\sol{f}\after e)}$:
\begin{equation}
\label{eq:2.2}
\vcenter{
\xy
\xymatrix{
&
X
\ar[0,2]^-{r}
\ar[1,0]_{e}
\ar `l[dl] `[dd]_{\sol{f} \after e} [dd]
&
&
A
\\
&
HX+Y
\ar[1,0]_{HX+\sol{f}}
&
&
\\
&
HX+A
\ar[0,2]_-{Hr+A}
&
&
HA+A
\ar[-2,0]_{[\alpha,A]}
}
\endxy
}
\end{equation}
We now only need to show that the passages from $HX+Y$ to $A$
in the above squares~\refeq{eq:2.1} and~\refeq{eq:2.2}
are equal. The left-hand components are, in both cases, 
$\alpha\cdot Hr:HX\to A$. For the right-hand components use 
$\sol{f}=[\alpha,A]\cdot (H\sol{f}+A)\cdot f$.
\end{proof}
\fi

\begin{rem}
  The same holds for every iterative algebra, except that here we
  work in a locally finitely presentable category and restrict $X$
  and $Y$ in Definition~\ref{dfn:cia} to finitely presentable objects. 
\end{rem}

\begin{rem}
  As mentioned in the Introduction, our two axioms, functoriality and
  compositionality, are not new as ideas of axiomatizing recursion---we believe
  however, that their concrete form is new, and their motivation strengthened
  by the results below. 
  
  Functoriality resembles the ``functorial dagger implication''
  of S.~Bloom and Z.~\'Esik~\cite{be}, 5.3.3, which states that for every
  object $p$ of an iterative theory the formation $f \mapsto \sol{f}$ of
  solutions for ideal morphisms $f: m \to m+p$ is a functor.
  Compositionality resembles the ``left pairing identity'' of~\cite{be}, 5.3.1,
  which for $f: n \to n + m + p$ and $g: m \to n + m + p$ states that
  $$
  \sol{[f,g]} = [\sol{f} \cdot [\sol h, \id_p], \sol h]\,,
  $$
  where
  $$
  h \equiv
  \xymatrix{
    m \ar[r]^-g 
    &
    n + m + p 
    \ar[rr]^-{[\sol{f}, \id_{m+p}]}
    & &
    m + p\,.
    }
  $$
  This identity corresponds also to the Beki\'c-Scott identity, see
  e.\,g.~\cite{moss}, 2.1.
\end{rem}

\section{Elgot Algebras}
\label{sec:3}
\setcounter{equation}{0}

\begin{defi}
\label{def:3.1}
Let $H$ be an endofunctor of a category with finite coproducts.
An {\em Elgot algebra} is  an $H$-algebra
$\alpha:HA\to A$ together with a function $\sol{({-})}$
which to every finitary flat equation morphism 
$$
e:X\to HX+A
$$
assigns a solution $\sol{e}:X\to A$ in such a way that
functoriality~\refeq{eq:func} and compositionality~\refeq{eq:bs-identity}
are satisfied.

By a \emph{complete Elgot algebra} we analogously understand an $H$-algebra
together with a function $\solop$ assigning to every flat equation $e$ a
solution $\sol{e}$ so that functoriality and compositionality are satisfied. 
\end{defi}

\begin{exa}\label{ex:jsl}
  Every join semilattice $A$ is an Elgot algebra. More precisely: consider the
  polynomial endofunctor $HX = X \times X$ of $\Set$ (expressing one binary
  operation). Then for every join semilattice $A$ there is a ``canonical''
  Elgot algebra structure on $A$ obtained as follows: the algebra $RA$ of
  all rational binary trees on $A$ has an interpretation on $A$ given by the
  function $\alpha: RA \to A$ forming, for every rational binary tree $t$ the
  join of all the (finitely many!) labels of leaves of $t$ in $A$. Now given a
  finitary flat equation morphism $e: X \to X \times X + A$, it has a unique
  solution $\sol{e}: X \to RA$ in the free iterative algebra $RA$, and composed
  with $\alpha$ this yields an Elgot algebra structure
  $e \mapsto \alpha \cdot\sol{e}$
  on $A$. See Example~\ref{ex:jsl2} for a proof.
\end{exa}
\begin{rem}
  In contrast, no nontrivial join semilattice is iterative. In fact, in an
  iterative join semilattice there must be a unique solution of the formal
  equation $x \approx x \vee x$. 
\end{rem}

\begin{exa}\label{ex:cpo}
  Continuous algebras on cpos are complete Elgot algebras. 
Let us work here in the category 
$$
\CPO
$$
of all {\em $\omega$-complete posets},
which are posets having joins of increasing
$\omega$-chains; morphisms are the {\em continuous functions},
i.e., functions preserving joins of $\omega$-chains. 
A functor $H:\CPO\to\CPO$ is called {\em locally continuous} provided
that for arbitrary CPOs, $X$ and $Y$, the associated function from
$\CPO(X,Y)$ to $\CPO(HX,HY)$ is continuous (i.e., 
$H(\bigsqcup f_n)=\bigsqcup Hf_n$ holds for all increasing 
$\omega$-chains $f_n:X\to Y$). For example, every polynomial
endofunctor $X\mapsto\coprod_n \Sigma_n\times X^n$ of $\CPO$ (where $\Sigma_n$
are cpos) is locally continuous.

Observe that the category $\CPO$ has coproducts: they are the disjoint unions 
with elements of different summands incompatible.

\begin{prop}\label{prop:cpo}
  Let $H:\CPO\to\CPO$ be a locally continuous functor and let $\alpha:HA\to A$
  be an $H$-algebra with a least element $\bot \in A$.  Then $(A, \alpha,
  \solop)$ is a complete Elgot algebra w.r.t.~the assignment of the least
  solution $\sol e$ to every flat equation morphism $e$. 
\end{prop}
\iffull
\begin{rem}\label{rem:cpo}
\fi
Notice that the least solution of $e:X\to HX+A$ refers to the elementwise order
of the hom-set $\CPO(X,A)$. We can actually prove a concrete 
formula for $\sol{e}$ as a join of the $\omega$-chain
$$
\sol{e}=\bigsqcup_{n\in\omega}\sol{e}_n
$$
of ``approximations'': $\sol{e}_0$
is the constant function to $\bot$, the least element of $A$,
and given $\sol{e}_n$, then $\sol{e}_{n+1}$ is defined by the commutativity
of~\refeq{eq:approxsol}. 
\iffull
\end{rem}
\fi
\iffull
\begin{proof}[Proof of Proposition~\ref{prop:cpo}]
  (1)~Let $e:X\to HX+A$ be a flat equation morphism in $A$.
  We define a function $F$ on $\CPO(X,A)$ by
  $$
  F:(s:X\to A)\mapsto ([\alpha,A]\cdot (Hs+A)\cdot e).
  $$
  Since $H$ is locally continuous and composition in the
  category $\CPO$ is continuous, we see that $F$ is continuous
  too.

  By the Kleene Fixed Point Theorem, there exists a least
  fixed point of $F$ and this is the least solution as described in
  Remark~\ref{rem:cpo}.  

\smallskip\noindent
(2)~The assignment $e\mapsto \sol{e}$ is functorial. In fact, let
      $$
      \xy
      \xymatrix{
      X
      \ar[0,2]^-{e}
      \ar[1,0]_{h}
      &
      &
      HX+A
      \ar[1,0]^{Hh+A}
      \\
      Y
      \ar[0,2]_-{f}
      &
      &
      HY+A
      }
      \endxy
      $$
      be a coalgebra homomorphism. It is easy to see by induction that
      $$
      \sol{e}_n=\sol{f}_n\cdot h
      \quad
      \mbox{(for all $n\geq 0$)}\,,
      $$
      thus, $\sol{e}=\sol{f}\cdot h$.

\smallskip\noindent
(3)~We prove compositionality. Let
      $$
      e:X\to HX+Y
      \quad
      \mbox{and}
      \quad
      f:Y\to HY+A
      $$
      be given. We shall show that the equality
      $$
      \sol{(f\plus e)}\cdot\inl=\sol{(\sol{f}\after e)}
      $$
      holds. It suffices to prove, by induction on $n$, that the following two
      inequalities 
      \begin{eqnarray}
        \sol{(f\plus e)}_n\cdot\inl
        &\less&
        \sol{(\sol{f}\after e)}
        \label{eq:ind1} 
        \\
        \sol{(\sol{f}\after e)}_n
        &\less&
        \sol{(f\plus e)}\cdot\inl
        \label{eq:ind2}
      \end{eqnarray}
      hold.
      First recall that $\inr:(Y,f)\to (X+Y,f\plus e)$ is a coalgebra
      homomorphism. Thus, the equation
      $
      \sol{(f\plus e)}\cdot\inr=\sol{f}
      $
      holds by functoriality.
      For the induction step for~\refeq{eq:ind1} consider the
      following diagram  
      \begin{displaymath}
      \vcenter{
      \xy
      \xymatrix@R-4pt{
      X
      \ar[0,4]^-{\sol{(\sol{f}\after e)}}
      \ar[2,0]_{e}
      \ar[1,2]_{\inl}
      &
      &
      &
      &
      A
      \\
      &
      &
      X+A
      \ar[2,0]^{f\smallplus e}
      \ar[-1,2]_{\sol{(f\smallplus e)}_{n+1}}
      \ar @{} [-1,0]|{\begin{turn}{90}$\labelstyle\less$\end{turn}}
      &
      &
      \\
      HX+Y
      \ar[3,0]_{HX+\sol{f}}
      \ar[1,1]^{HX+f}
      \ar @{} [0,2]|-{=} 
      &
      &
      \ar @{} [0,2]|-{=} 
      &
      &
      \\
      &
      HX+HY+A
      \ar[0,1]^-{\can+A}
      \ar[2,3]_*+{\labelstyle [H\sol{(\sol{f}\after e)},H\sol{f}]+A}
      &
      H(X+Y)+A
      \ar[2,2]^{H\sol{(f\smallplus e)}_n+A}
      \ar @{} [1,0]|{\begin{turn}{45}$\labelstyle\sqsupseteq$\end{turn}}
      &
      &
      \\
      &
      &
      &
      &
      \\
      HX+A
      \ar[0,4]_-{H\sol{(\sol{f}\after e)}+A}
      &
      &
      &
      &
      HA+A
      \ar[-5,0]_{[\alpha,A]}
      }
      \endxy
      }
      \end{displaymath}
      In order to prove the desired inequality in the upper triangle, we use
      the fact that the outer square commutes by definition of $\solop$.
      The three middle parts clearly behave as indicated (for the triangle use
      the induction hypothesis~\refeq{eq:ind1} and $\sol{(f \plus e)}_n \cdot \inr
      \less \sol{(f \plus e)} \cdot \inr = \sol{f}$), and the lowest part commutes
      when extended by $[\alpha,A]$: In
      fact, for the left-hand component with domain $HX$ this is trivial; for
      the right-hand component with domain $Y$ use $\sol{f} = [\alpha, A]\cdot
      (H\sol f +A) \cdot f$, see~\refeq{diag:sol}.

      For the induction step for~\refeq{eq:ind2} consider the following diagram
      \begin{displaymath}
        \vcenter{
          \xy
          \xymatrix{
            X
            \ar[0,4]^-{\sol{(\sol{f}\after e)}_{n+1}}
            \ar[2,0]_{e}
            \ar[1,2]_{\inl}
            &
            &
            &
            &
            A
            \\
            &
            &
            X+A
            \ar[2,0]^{f\smallplus e}
            \ar[-1,2]_{\sol{(f\smallplus e)}}
            \ar @{} [-1,0]|{\begin{turn}{90}$\labelstyle\sqsupseteq$\end{turn}}
            &
            &
            \\
            HX+Y
            \ar[3,0]_{HX+\sol{f}}
            \ar[1,1]^{HX+f}
            \ar @{} [0,2]|-{=} 
            &
            &
            \ar @{} [0,2]|-{=} 
            &
            &
            \\
            &
            HX+HY+A
            \ar[0,1]^-{\can+A}
            \ar[2,3]_*+{\labelstyle [H\sol{(\sol{f}\after e)}_n,H\sol{f}]+A}
            &
            H(X+Y)+A
            \ar[2,2]^{H\sol{(f\smallplus e)} + A}
            \ar @{} [1,0]|{\begin{turn}{45}$\labelstyle\less$\end{turn}}
            &
            &
            \\
            &
            &
            &
            &
            \\
            HX+A
            \ar[0,4]_-{H\sol{(\sol{f}\after e)}_n+A}
            &
            &
            &
            &
            HA+A
            \ar[-5,0]_{[\alpha,A]}
            }
          \endxy
          }
      \end{displaymath}
      
      The outer square commutes by definition of $\sol{(\sol{f} \after
        e)}_{n+1}$. The three middle parts behave as indicated (for the
      inequality use the induction hypothesis), and the lowest part commutes
      when extended by $[\alpha, A]$ as before. Thus, we obtain the desired
      inequality in the upper triangle.
\end{proof}
\fi
\end{exa}
\begin{rem}
\label{rem:cpoelgot}
Many set functors $H$ have a lifting to locally continuous
endofunctors $H'$ of $\CPO$. That is, for the forgetful functor
$U:\CPO\to\Set$ the following square
$$
\xy
\xymatrix{
\CPO
\ar[0,2]^-{H'}
\ar[1,0]_{U}
&
&
\CPO
\ar[1,0]^{U}
\\
\Set
\ar[0,2]_-{H}
&
&
\Set
}
\endxy
$$
commutes. For example, every polynomial functor $H_\Sigma$ has
such a lifting. An $H$-algebra \mbox{$\alpha:HA\to A$} is called  
{\em $\CPO$-enrichable} if there exists a CPO-ordering $\less$ 
with a least element on the set $A$ such that $\alpha$ is a continuous
function from $H'(A,\less)$ to $(A,\less)$. 

\begin{cor}\label{cor:cpo}
  Every $\CPO$-enrichable $H$-algebra $A$ in $\Set$ is a complete
  Elgot algebra. 
\end{cor}
In fact, to every equation morphism $e:X\to HX+A$ assign the 
least solution of $e:(X,\leq)\to {H'(X,\leq)}+(A,\less)$ where $\leq$
is the discrete ordering of $X$ ($x\leq y$ iff $x=y$).
\end{rem}

\comment{
\begin{exa}
  Corollary~\ref{cor:cpo} does not cover all complete Elgot algebras---even for
  simple functors such as $H = \Id + \Id$ (i.\,e., even for algebras with 
  two unary operations). That is, there exist complete Elgot algebras which
  are not $\CPO$-enrichable. This was demonstrated by J.~Tiuryn, see 6.15
  of~\cite{t}: 

  Let $A$ be the unit interval $[0,1]$ with two unary operations
  $$
  f(x) = \frac{1}{2}x
  \qquad
  \textrm{and}
  \qquad
  g(x) = \frac{1}{2}x + \frac{1}{2}\,.
  $$
  This algebra is a CIA by Corollary~\ref{cor:metrizable}: the usual metric is
  complete and $f$ and $g$ are contracting maps. However, there does not exist
  a $\CPO$ structure on $A$ making $f$ and $g$ continuous. 
\end{exa}
}

\begin{exa}
\label{ex:3.2}
\emph{Unary algebras.}  Let $H=\Id$ as an endofunctor of $\Set$. Given an
$H$-algebra $ \alpha:A\to A\,, $ if $\alpha$ has no fixed point, then $A$
carries no Elgot algebra structure: consider the equation $x \approx
\alpha(x)$.

Conversely, every fixed point $a_0$ of $\alpha$ yields a flat cpo structure
with a least element $a_0$ on $A$, i.\,e., $x \leq y$ iff $x = y$ or $x =
a_0$. Thus, $A$ is a complete Elgot algebra since it is CPO-enrichable.
Notice that for every flat equation morphism $e: X \to X + A$, the least
solutions $\sol{e}$ operates as follows: for a variable $x$ we have
$$
\sol{e}(x) = \left\{
  \begin{array}{c@{\qquad}p{9cm}}
    \alpha^k(a) & if there is a sequence $x = x_0, x_1, \ldots x_k$ in $X$ that
    fulfils $e(x_0) = x_1, \ldots e(x_{k-1}) = x_k$ and $e(x_k) = a$ 
    \\
    a_0 & else\,.
  \end{array}
\right.
$$
\end{exa}

\iffull
\begin{rem}
  For unary algebras, Example~\ref{ex:3.2} describes \emph{all} existing Elgot
  algebras. In fact, let $(A, \alpha, \solop)$ be an Elgot algebra and let
  $a_0\in A$ be the chosen solution of $x \approx \alpha(x)$;
  more precisely, $a_0=\sol{e}(*)$ where $e=\inl: 1\to 1 + A$
  and $*$ is the unique element of $1$.
  Then for every flat equation morphism $e: X \to X + A$ the
  chosen solution sends a variable $x \in X$ to one of the above values
  $\alpha^k(a)$ or $a_0$. To prove this denote by $Y \subseteq X$ the set of
  all variables for which the ``else'' case holds above.
  Hence no sequence $x = x_0, \ldots x_k$ in $X$ fulfils
  $e(x_i) = x_{i+1}$, for $i = 0, \ldots, k-1$,
  and $e(x_k) \in A$. Apply functoriality to the morphism $h$ from $e$
  to $1+e: 1+X \to 1+X + A$ defined by $h(y) \in 1$ for $y \in Y$ and $h(x) = x
  \in X$ else. In fact, the chosen solution of the unique element of $1$ in
  $1+X$ must be $a_0$ by functoriality (consider the left-hand coproduct
  injection from the flat equation morphism $\inl: 1 \to 1 + A$ to $1 + e$).
\end{rem}
\fi

\iffull
\begin{exa}
  Just as Example~\ref{ex:cpo} is based on the Kleene Fixed Point Theorem, we
  obtain examples of complete Elgot algebras based on the Knaster-Tarski Fixed
  Point Theorem. Here we work with the category
  $$
  \Pos
  $$
  of all posets and order-preserving functions. (In fact, everything we say
  holds, much more generally, in every category enriched over $\Pos$ with
  $\Pos$-enriched finite coproducts.) A functor $H: \Pos \to \Pos$ is called 
  \emph{locally order-preserving} if for all order-preserving functions $f, g:
  A \to B$ with $f \sqsubseteq g$ (in the pointwise ordering of $\Pos(A,B)$, of
  course), we have $Hf \sqsubseteq Hg$.
\end{exa}

\begin{prop}\label{prop:pos}
  Let $H: \Pos \to \Pos$ be locally order-preserving and let $\alpha: HA \to A$
  be an $H$-algebra which is carried by a complete lattice $A$. Then $(A,
  \alpha, \solop)$ is a complete Elgot algebra w.r.t.~the assignment of a least
  solution $\sol{e}$ to every flat equation morphism $e$.
\end{prop}

\begin{rem}
  Again, the least solution of $e:X\to HX+A$ refers to the elementwise order
  of the hom-set $\Pos(X,A)$. We can actually prove a concrete 
  formula for $\sol{e}$ as a join of the ordinal chain
  $$
  \sol{e}=\bigsqcup_{n\in\Ord}\sol{e}_n
  $$
  of ``approximations'': $\sol{e}_0$
  is the constant function to $\bot$, the least element of $A$,
  given $\sol{e}_n$, then $\sol{e}_{n+1}$ is defined by the commutativity
  of~\refeq{eq:approxsol} and for limit ordinals $n$ we put $\sol{e}_n =
  \bigsqcup_{k < n} \sol{e}_k$. 
\end{rem}
\begin{proof}[Proof of Proposition~\ref{prop:pos}]
  One essentially repeats the proof of Proposition~\ref{prop:cpo} for $\sol{e}$
  as defined in the previous Remark. In part~(1) apply the Knaster-Tarski Fixed
  Point Theorem in lieu of the Kleene Fixed Point Theorem. For part~(2) replace
  every induction argument by a corresponding transfinite induction argument
  and notice that the limit step is always trivial.  
\end{proof}
\fi

\begin{exa}\label{ex:lat}
  Every complete lattice $A$ is a complete Elgot algebra for $HX = X \times X$.
  Analogously to Example~\ref{ex:jsl} we have a function $\alpha: TA \to A$
  assigning to every binary tree $t$ in $TA$ the join of all labels of leaves
  of $t$ in $A$. Now for every flat equation morphism $e$ in $A$ we have its
  unique solution $\sol{e}$ in $TA$ and this yields a
  complete Elgot algebra structure $e \mapsto \alpha \cdot \sol{e}$. See
  Example~\ref{ex:prooflattice} for a proof.
\end{exa}

\section{The Eilenberg-Moore Category of the Monad $\R$}
\label{sec:4}
\setcounter{equation}{0}

We prove now that the category of all Elgot algebras and solution-preserving
morphisms, defined as expected, is the category $\kat{A}^\R$ of
Eilenberg-Moore algebras of the rational monad $\R$ of $H$, see
Remark~\ref{rem:rat}. 

\begin{asm}
Throughout this section $H$ denotes a finitary endofunctor
of a locally finitely presentable category $\kat{A}$.
\end{asm}

We denote
by $\kat{A}_\f$ a small full subcategory representing all finitely
presentable objects of $\kat{A}$. Recall the operations $\after$ and $\plus$ 
from Remark~\ref{rem:2.7}.

\begin{defi}
  \label{dfn:solpres}
  Let $(A,\alpha,\sol{({-})})$, and $(B,\beta,\Sol{({-})})$ be Elgot
  algebras.  We say that a morphism $h:A\to B$ in $\kat{A}$ {\em preserves
    solutions} provided that for every finitary flat equation morphism
  \mbox{$e:X\to HX+A$} we have the following equation 
  \begin{equation}\label{eq:solpres}
    \xymatrix@1{
      X \ar[r]^-{\sol{e}} & A \ar[r]^-h & B
      }
    \equiv
    \xymatrix@1{
      X \ar[rr]^-{\Sol{(h\after e)}} & & B\,.
      }
  \end{equation}
\end{defi}

\begin{lem}
\label{lem:solpres}
Every solution-preserving morphism between Elgot algebras is a homomorphism of
$H$-algebras, i.e., we have $h \cdot \alpha = \beta \cdot Hh$. 
\end{lem}
\iffull
\begin{proof}
  Let $\kat{A}_\f /A$ be the comma-category of all arrows $q: X
  \to A$ with $X$ in $\kat{A}_\f$. Since $\kat{A}$ is locally finitely
  presentable, $A$ is a filtered colimit of the canonical diagram
  $D_A: \kat{A}_\f /A\to\kat{A}$ given by $(q:X\to A)\mapsto X$.

  Now $\kat{A}_\f$ is a generator of $\kat{A}$, thus, in order to prove the
  lemma it is sufficient to prove 
  that for every morphism $p:Z\to HA$ with $Z$ in $\kat{A}_\f$ we have
  \begin{equation}
    \label{eq:p}
    h\cdot\alpha\cdot p=\beta\cdot Hh\cdot p.
  \end{equation}
  
  Since $H$ is finitary, it preserves the above colimit $D_A$. This implies,
  since $\kat{A}(Z, -)$ preserves filtered colimits, that $p$ has a
  factorization 
  $$
  \xy
  \xymatrix{
    Z
    \ar[0,2]^-{p}
    \ar[1,2]_{s}
    &
    &
    HA
    \\
    &
    &
    HX
    \ar[-1,0]_{Hq}
    }
  \endxy
  $$
  for some $q: X\to A$ in $\kat{A}_\f/ A$ and some $s$. For the
  following equation morphism
  $$
  \xy
  \xymatrix@C+1pc{
    e\equiv Z+X
    \ar[r]^-{s+X}
    &
    HX+X
    \ar[r]^-{H\inr+q}
    &
    H(Z+X)+A
    }
  \endxy
  $$
  we have a commutative square 
  $$
  \xy
  \xymatrix{
    & Z+X
    \ar[0,2]^-{\sol{e}}
    \ar[1,0]_{s+X}
    \ar `l[ld] `[dd]_e [dd] 
    &
    &
    A
    \\
    & HX+X
    \ar[1,0]_{H\inr+q}
    &
    &
    \\
    & H(Z+X)+A
    \ar[0,2]_-{H\sol{e}+A}
    &
    &
    HA+A
    \ar[-2,0]_{[\alpha,A]}
    }
  \endxy
  $$
  Consequently, $\sol{e} \cdot \inr = q$, and this implies $\sol{e} \cdot \inl =
  \alpha \cdot H(\sol{e}\cdot \inr) \cdot s = \alpha \cdot p$. 
  Since $h$ preserves solutions, we have 
  $h\cdot\sol{e}=\Sol{(h\after e)}$ and therefore
  \begin{equation}
    \label{eq:pp}
    \Sol{(h\after e)}= [h\cdot \alpha\cdot p,h\cdot q].
  \end{equation}
  On the other hand, consider the following diagram
  $$
  \xy
  \xymatrix{
    & Z+X
    \ar[0,4]^-{\Sol{(h\after e)}}
    \ar[1,0]_{s+X}
    \ar[1,3]^{p+hq}
    \ar `l[ld] `[ddd]_{h \after e} [ddd]
    &
    &
    &
    &
    B
    \\
    & HX+X
    \ar[1,0]_{H\inr+q}
    \ar[0,3]^-{Hq+hq}
    &
    &
    &
    HA+B
    \ar[2,1]^(.4){Hh+B}
    &
    \\
    & H(Z+X)+A
    \ar[1,0]_{H(Z+X)+h}
    \ar[-1,3]^(.4)*+{\labelstyle H[\alpha p,q]+h}
    &
    &
    &
    &
    \\
    & H(Z+X)+B
    \ar[0,4]_-{H\Sol{(h\after e)}+B}
    \ar[-2,3]_*+{\labelstyle H[\alpha p,q]+B}
    &
    &
    &
    &
    HB+B
    \ar[-3,0]_{[\beta,B]}
    }
  \endxy
  $$
  It commutes: the outer shape commutes since $\Sol{(h\after e)}$ is a
  solution. For the lower triangle use equation~\refeq{eq:pp}, and the
  remaining triangles are trivial. Thus, the upper right-hand part commutes:
  \begin{equation}
    \label{eq:ppp}
    \Sol{(h\after e)}=[\beta\cdot Hh\cdot p,h\cdot q].
  \end{equation}
  Now the left-hand components of~\refeq{eq:pp} and~\refeq{eq:ppp}
  establish the desired equality~\refeq{eq:p}.
\end{proof}
\fi
\begin{exa}
  The converse of Lemma~\ref{lem:solpres} is true for iterative algebras, as
  proved in~\cite{amv}, but for Elgot algebras in general it is false. In
  fact, consider the unary algebra $\id:A\to A$, where $A=\{\, 0,1\,\}$.  This
  is an Elgot algebra with the solution structure $\sol{({-})}$ given by the
  fixed point $0\in A$, see Example~\ref{ex:3.2}.

Then ${\mathrm{const}}_1:A\to A$ is a homomorphism of unary algebras
that does not preserve solutions. Indeed, consider the following 
equation morphism 
$$
e:\{ x\}\to\{ x\}+A,
\quad
x\mapsto x.
$$
We have $\sol{e}(x)=0$, and thus
$
1
=
{\mathrm{const}}_1\cdot\sol{e}(x)
\not=
\sol{({\mathrm{const}}_1\after e)}(x)
=
\sol{e}(x) 
= 
0
$.
\end{exa}

\begin{notation}
We denote by
$$
\Elgot{H}
$$
the category of all Elgot algebras and solution-preserving morphisms.
\end{notation}

\iffull\else
\begin{prop}
\label{prop:4.7}
A free iterative algebra on $Y$ is a free Elgot algebra
on $Y$.
\end{prop}
\fi

\begin{rem}\label{rem:prop}
  For the two operations $\after$ and $\plus$ from Remark~\ref{rem:2.7} we list
  some obvious properties that these operations have for all $e: X \to HX + Y$,
  $f: Y \to HY + Z$, $s: Z \to Z'$ and $t: Z' \to Z''$: 
  \begin{enumerate}
  \item\label{item:2.7.1} $\id_Y\after e=e$. \iffull This is trivial. \fi

  \item\label{item:2.7.2} $t\after (s\after e)=(t\cdot s)\after e$.
    \iffull
    \hfill\\ 
      See the following diagram
      $$
      \xy
      \xymatrix{
      X
      \ar[0,2]^-{e}
      &
      &
      HX+Y
      \ar[0,2]^-{HX+s}
      \ar[1,2]_{HX+t\cdot s}
      &
      &
      HX+Y'
      \ar[1,0]^{HX+t}
      \\
      &
      &
      &
      &
      HX+Y''
      }
      \endxy
      $$ 
      \fi
\item\label{item:2.7.4} $s\after (f\plus e)=(s\after f)\plus e$.
  \iffull
  \hfill\\
      See the following diagram 
      $$
      \xy
      \xymatrix{
      X+Y
      \ar[0,1]^-{[e,\inr]}
      &
      HX+Y
      \ar[0,1]^-{HX+f}
      \ar[1,0]_{HX+s\after f}
      &
      HX+HY+Z
      \ar[0,1]^-{\can+Z}
      &
      H(X+Y)+Z
      \ar[1,0]_{H(X+Y)+s}
      \\
      &
      HX+HY+Z'
      \ar[0,2]_-{\can +Z'}
      &
      &
      H(X+Y)+Z'      
      }
      \endxy
      $$
      \fi
\end{enumerate} 
\end{rem}

\iffull
\begin{prop}
\label{prop:4.7}
A free iterative algebra on $Y$ is a free Elgot algebra
on $Y$.
\end{prop}
\begin{proof}
(1)~ We first recall the construction of the free iterative algebra $RY$ on
$Y$ presented in~\cite{amv}. For the functor $H(-) + Y$ denote by
$$
\EQ{Y}
$$
the full subcategory of $\Coalg{(H(-) + Y)}$ given by all coalgebras with
a finitely presentable carrier, i.e., finitary flat equation morphisms
$e:X\to HX+Y$. The inclusion functor
$\Eq{Y}: \EQ{Y}\to\Coalg{(H(-) + Y)}$ is an essentially small filtered
diagram. Put  
$$
RY = \colim \Eq{Y}\,.
$$
More precisely, form a colimit of the above diagram $\Eq{Y}$. This is a
coalgebra $RY$ with the following coalgebra structure
$$
i:RY\to HRY+Y
$$
and with colimit injections
$$
\inj{e}:(X,e)\to (RY,i)\qquad \textrm{for all $e: X \to HX + Y$ in $\EQ{Y}$.}
$$
Notice that this colimit is preserved 
by the forgetful functor $\Coalg{(H(-) +
  Y)} \to \kat{A}$ since $H$ is finitary.  

The coalgebra structure $i:RY\to HRY+Y$ is an isomorphism; its inverse gives an
$H$-algebra structure
$$
\rho_Y:HRY\to RY
$$
and a morphism
$$
\eta_Y:Y\to RY.
$$
Furthermore, we proved that the algebra $(RY,\rho_Y)$ is a free iterative $H$-algebra on
$Y$ with the universal arrow $\eta_Y$.

Recall further from~\cite{amv} that the unique solution 
$$
\Sol{e}:X\to RY
$$
for a finitary flat equation morphism $e:X\to HX+RY$
is obtained as follows. There exists a factorization
\begin{equation}
\label{eq:4.1}
\vcenter{
\xy
\xymatrix{
X
\ar[0,2]^-{e}
\ar[1,2]_{e_0}
&
&
HX+RY
\\
&
&
HX+Z
\ar[-1,0]_{HX+\inj{g}}
}
\endxy
}
\end{equation}
with $g: Z \to HZ+Y$ in $\EQ{Y}$. Define 
$$
\xy
\xymatrixcolsep{3pc}
\xymatrix{
\Sol{e}\equiv X
\ar[0,1]^-{\inl}
&
X+Z
\ar[0,1]^-{\inj{(g\smallplus e_0)}}
&
RY
}
\endxy
$$
This defines $\Sol{({-})}$ from $\inj{({-})}$. Conversely, it
is not difficult to see that the equality
\begin{equation}\label{eq:inj_sol}
  \inj{e} =\Sol{(\eta_Y\after e)}
\end{equation}
holds for every $e:X\to HX+Y$ in $\EQ{Y}$. Finally, the universal arrow
$\eta_Y$ has for any finitely presentable object $Y$ the form
$\eta_Y=\inj{\inr}$ (for $\inr:Y\to HY+Y$).

\medskip\noindent 
(2)~We are prepared to prove the Proposition. Suppose that
$(A,\alpha,\solop)$ is an Elgot algebra and let $m:Y\to A$ be a morphism. We
are to prove that there exists a unique solution-preserving $h:RY\to A$ with
$h\cdot\eta_Y=m$.

In order to show existence, we define a morphism $h:RY\to A$
by commutativity of the following triangles
$$
\xy
\xymatrix{
RY
\ar[0,2]^-{h}
&
&
A
\\
X
\ar[-1,0]^{\inj{e}}
\ar[-1,2]_{\sol{(m\after e)}}
}
\endxy
$$
for all $e:X\to HX+Y$ in $\EQ{Y}$. The definition of $h$ is justified,
since the morphisms $\sol{(m\after e)}$ form a cocone for $\Eq{Y}$:
for any coalgebra homomorphism 
$
k:(X,e)\to (Z,g)
$
in $\EQ{Y}$ we have a coalgebra homomorphism
$
k:(X,m\after e)\to (Z,m\after g).
$
Thus, 
$
\sol{(m\after e)}\cdot k=\sol{(m\after g)}
$
holds by functoriality.
For $e=\inr:Y\to HY+Y$,
$Y$ finitely presentable, we have
$\inj{e}=\eta_Y$, thus, 
$$
\begin{array}{rcl@{\qquad}l}
  h\cdot\eta_Y & = & \sol{(m\after\inr)} & \mbox{(since $\eta_Y = \inj\inr$)} \\
  & = & [\alpha, A] \cdot (H\sol{(m\after\inr)} + A) \cdot (m \after \inr) & 
  \mbox{(by~\refeq{diag:sol})} \\
  & = & [\alpha, A] \cdot (H\sol{(m\after\inr)} + A) \cdot (HY + m) \cdot \inr
  & \mbox{(Definition of $\after$)} \\
  & = & m\,.
\end{array} 
$$
For arbitrary objects $Y$ the equation $h\cdot\eta_Y=m$ follows easily.

Let us show that $h$ preserves solutions. 
We have
\def\macr{\\ & &}
$$
\begin{array}{rcl@{\qquad}l}
h\cdot\Sol{e} &=& h\cdot \inj{(g\plus e_0)}\cdot\inl &\mbox{(Definition of $\Sol{e}$)}\\
&=& \sol{(m\after (g\plus e_0))}\cdot\inl &\mbox{(Definition of $h$)}\\
&=& \sol{((m\after g)\plus e_0)}\cdot\inl &
    \mbox{(\ref{rem:prop}(\ref{item:2.7.4}))}\\
&=& \sol{(\sol{(m\after g)}\after e_0)} &\mbox{(compositionality)}\\
&=& \sol{((h\cdot\inj{g})\after e_0)} &\mbox{(Definition of $h$)}\\
&=& \sol{(h\after (\inj{g}\after e_0))} &
    \mbox{(\ref{rem:prop}(\ref{item:2.7.2}))}\\
&=& \sol{(h\after e)}
  & 
  \mbox{(\refeq{eq:4.1} and the definition of $\after$)}
\end{array}
$$

Concerning uniqueness, suppose that $h$ with $h\cdot\eta_Y = m$
preserves solutions, then we have 
$$
\begin{array}{rcl@{\qquad}l}
h\cdot \inj{e} 
&=& h\cdot \Sol{(\eta_Y\after e)} 
  & \mbox{(by~\refeq{eq:inj_sol})}\\
&=& \sol{(h\after(\eta_Y\after h))} 
  & \mbox{($h$ preserves solutions)} \\
&=& \sol{((h\cdot\eta_Y)\after e)} &
    \mbox{(\ref{rem:prop}(\ref{item:2.7.2}))}\\
&=& \sol{(m\after e)}
  & \mbox{(since $h\cdot\eta_Y=m$)}
\end{array}
$$
which determines $h$ uniquely.
\end{proof}
\fi
\begin{thm}
\label{th:4.6}
The category $\Elgot{H}$ of Elgot algebras is isomorphic 
to the Eilenberg-Moore category $\kat{A}^\R$ of $\R$-algebras
for the rational monad $\R$ of $H$.
\end{thm}
\begin{rem}
  The shortest proof we know is based on Beck's Theorem, see below. But
  this proof is
  not very intuitive. A slightly more technical (and much more illuminating)
  proof has the following sketch: Denote for any object $Y$ by $(RY, \rho_Y,
  \Solop)$ a free Elgot algebra on $Y$ with a universal arrow $\eta_Y: Y \to
  RY$.
\begin{enumerate}
\item For every $\R$-algebra $\alpha_0:RA\to A$ we have an
  ``underlying'' $H$-algebra
  $$
  \xy
  \xymatrix{
    \alpha\equiv HA
    \ar[r]^-{H\eta_A}
    & 
    HRA
    \ar[r]^-{\rho_A}
    &
    RA
    \ar[r]^-{\alpha_0}
    &
    A,
    }
  \endxy
  $$
  and the following formula for solving equations: given a finitary flat
  equation morphism $e:X\to HX+A$
  put
  $$
  \xy
  \xymatrix{
    \sol{e}\equiv X
    \ar[rr]^-{\Sol{(\eta_A \after e)}}
    &
    &
    RA
    \ar[0,1]^-{\alpha_0}
    &
    A\,.
    }
  \endxy
  $$
  It is not difficult to see that this formula indeed yields a choice of
  solutions satisfying functoriality and compositionality.
\item Conversely, given an Elgot algebra $\alpha:HA\to A$, define
  $\alpha_0:RA\to A$ as the unique solution-preserving morphism such that
  $\alpha_0 \cdot \eta_A = \id$. It is easy to see that $\alpha_0$ satisfies the
  two axioms of an Eilenberg-Moore algebra.
\item It is necessary to prove that the above passages extend to the level of
  morphisms and they form functors which are inverse to each other.
\end{enumerate}
\end{rem}
\begin{proof}(Theorem~\ref{th:4.6}.)
By Proposition~\ref{prop:4.7} the natural forgetful functor
$U:\Elgot{H}\to\kat{A}$ has a left adjoint $Y\mapsto RY$. Thus,
the monad obtained by this adjunction is $\R$. We prove that
the comparison functor $K:\Elgot{H}\to\kat{A}^\R$ is an isomorphism,
using Beck's theorem (see~\cite{maclane}, Theorem~1 in
Section~VI.7).
Thus, we must prove that $U$ creates coequalizers of $U$-split
pairs. Let $(A,\alpha,\sol{({-})})$ and $(B,\beta,\Sol{({-})})$ 
be Elgot algebras, and let $f,g:A\to B$ be solution-preserving
morphisms with a splitting
$$
\xy
\xymatrix{
A
\ar @<.7ex> [0,2]^-{f}
\ar @<-.7ex> [0,2]_-{g}
&
&
B
\ar[0,2]^-{c}
\ar @<1ex> @/^1pc/ [0,-2]^-{t}
&
&
C
\ar @/^1pc/ [0,-2]^-{s}
}
\endxy
$$
in $\kat{A}$ (where $cs=\id$, $ft=\id$ and $gt=sc$). Since $c$
is, then, an absolute coequalizer of $f$ and $g$, $c$ is a coequalizer 
in $\Alg{H}$ for a unique $H$-algebra structure $\gamma:HC\to C$.
In fact, the forgetful functor $\Alg{H}\to\kat{A}$ creates every
colimit that $H$ preserves.

It remains to show that $C$ has a unique Elgot algebra structure
such that
\begin{enumerate}
  \renewcommand{\labelenumi}{(\arabic{enumi})}
\item $c$ preserves solutions, and
\item $c$ is a coequalizer in $\Elgot{H}$. 
\end{enumerate}
We establish~(1) and~(2) in several steps.

\medskip\noindent
(a)~An Elgot algebra on $(C,\gamma)$. For every finitary flat equation
morphism $e:X\to HX+C$ we prove that the following
morphism
           $$
           \xy
           \xymatrix{
           e^*\equiv X
           \ar[0,2]^-{\Sol{(s\after e)}}
           &
           &
           B
           \ar[0,2]^-{c}
           &
           &
           C
           }
           \endxy
           $$ 
           is a solution of $e$. In fact, the following diagram
           $$
           \xy
           \xymatrix@C+1pc{
           X
           \ar[rr]^-{\Sol{(s\after e)}}
           \ar[rd]^*+{\labelstyle s\after e}
           \ar[dd]_{e}
           &
           &
           B
           \ar[r]^-{c}
           &
           C
           \\
           &
           **[l] HX+B
           \ar[r]_-{H\Sol{(s\after e)} +B}
           &
           HB+B
           \ar[u]_{[\beta,B]}
           \ar[rd]^*+{\labelstyle Hc+c}
           &
           \\
           HX+C
           \ar[rrr]_-{H(c\cdot\Sol{(s\after e)})+C}
           \ar[ru]^*+{\labelstyle HX+s}
           &
           &
           &
           HC+C
           \ar[uu]_{[\gamma,C]}
           }
           \endxy
           $$
           clearly commutes.

           Functoriality: any coalgebra homomorphism
           $$
           \xy
           \xymatrix{
           X
           \ar[0,2]^-{e}
           \ar[1,0]_{h}
           &
           &
           HX+C
           \ar[1,0]^{Hh+C}
           \\
           Z
           \ar[0,2]_-{z}
           &
           &
           HZ+C
           }
           \endxy
           $$
           is, of course, a coalgebra homomorphism
           $
           h:(X,s\after e)\to (Z,s\after z)\,.
           $
           Thus, the equations
           $$
           e^*=c\cdot \Sol{(s\after e)}=c\cdot \Sol{(s\after z)}\cdot h = z^*\cdot h
           $$
           hold by functoriality of $\Sol{({-})}$.
           
           Let us prove compositionality: suppose we have 
           finitary flat equation morphisms
           $$
           e:X\to HX+Y
           \quad
           \mbox{and}
           \quad 
           k:Y\to HY+C
           $$
           Then we obtain the desired equation as follows:
           
           \setlongtables
           \begin{longtable}{>{$}r<{$}>{$}c<{$}>{$}l<{$}@{\hspace{1.5em}}l}
             \endhead
             \endfoot
           (k^*\after e)^* &=& c\cdot\Sol{(s\after (k^*\after e))} &
           \mbox{(Definition of $(-)^*$)} \\
           &=& c\cdot\Sol{(s\after (c\cdot\Sol{(s\after k)}\after e))} 
             & \mbox{(Definition of $(-)^*$)}\\ 
           &=& c\cdot\Sol{((s\cdot c)\after (\Sol{(s\after k)}\after e))} 
             & \mbox{(\ref{rem:prop}(\ref{item:2.7.2}))} \\
           &=& c\cdot\Sol{((g\cdot t)\after (\Sol{(s\after k)}\after e))} 
             & \mbox{($g\cdot t=s\cdot c$)}\\ 
           &=& c\cdot\Sol{(g\after (t\after (\Sol{(s\after k)}\after e)))} 
             & \mbox{(\ref{rem:prop}(\ref{item:2.7.2}))} \\
           &=& c\cdot g\cdot\sol{(t\after (\Sol{(s\after k)}\after e))}
             & \mbox{($g$ preserves solutions)} \\           
           &=& c\cdot f\cdot\sol{(t\after (\Sol{(s\after k)}\after e))}
             & \mbox{($c\cdot f=c\cdot g$)} \\ 
           &=& c\cdot\Sol{((f\cdot t)\after (\Sol{(s\after k)}\after e))} 
             & %
             \iffull %
             \mbox{%
               ($f$ preserves solutions
               and~\ref{rem:prop}(\ref{item:2.7.2}))
               } 
             \\%
             \else %
             \\ & & %
             \multicolumn{2}{l}{ %
               \mbox{
                 ($f$ preserves solutions
                 and~\ref{rem:prop}(\ref{item:2.7.2}))
                 } %
               } %
             \\
             \fi %
           &=& c\cdot\Sol{(\Sol{(s\after k)}\after e)}
             & \mbox{($f\cdot t=\id$
           and~\ref{rem:prop}\refeq{item:2.7.1})} \\
           &=& c\cdot \Sol{((s\after k)\plus e)}\cdot\inl
             & \mbox{(compositionality of $\Sol{({-})}$)}\\
           &=& c\cdot \Sol{(s\after (k\plus e))}\cdot\inl
             & %
             \iffull %
             \mbox{(Since $(s\after k) \plus e = s \after (k \plus e)$
               by~\ref{rem:prop}\refeq{item:2.7.4})} %
             \\%
             \else %
             \\ & & %
             \multicolumn{2}{l}{
               \mbox{(Since $(s\after k) \plus e = s \after (k \plus e)$
                 by~\ref{rem:prop}\refeq{item:2.7.4})}%
               } \\ %
             \fi %
           &=& (k\plus e)^*\cdot\inl & \textrm{(Definition of $(-)^*$)}           
         \end{longtable}
         
\medskip\noindent
(b)~The morphism $c:B\to C$ is solution-preserving. In fact,
for any finitary flat equation morphism
$
e:X\to HX+B
$
we have the desired equation:

           \setlongtables
           \begin{longtable}{>{$}r<{$}>{$}c<{$}>{$}l<{$}@{\hspace{2em}}l}
             \endhead
             \endfoot
           (c\after e)^* 
           &=& c\cdot \Sol{(s\after (c\after e))}
             & \mbox{(Definition of $({-})^*$)}\\
           &=& c\cdot \Sol{((s\cdot c)\after e)} 
             & \mbox{(\ref{rem:prop}(\ref{item:2.7.2}))} \\
           &=& c\cdot \Sol{((g\cdot t)\after e)} 
             & \mbox{($g\cdot t=s\cdot c$)}\\
           &=& c\cdot \Sol{(g\after (t\after e))} 
             & \mbox{(\ref{rem:prop}(\ref{item:2.7.2}))} \\
           &=& c\cdot g\cdot \sol{(t\after e)}
             & \mbox{($g$ preserves solutions)}\\
           &=& c\cdot f\cdot \sol{(t\after e)}
             & \mbox{($c\cdot f=c\cdot g$)}\\ 
           &=& c\cdot \Sol{(f\after (t\after e))} 
             & \mbox{($f$ preserves solutions)}\\
           &=& c\cdot \Sol{((f\cdot t)\after e)} 
             & \mbox{(\ref{rem:prop}(\ref{item:2.7.2}))}\\
           &=& c\cdot \Sol{(\id\after e)} 
             & \mbox{($f\cdot t=\id$)}\\
           &=& c\cdot\Sol{e} 
             & \mbox{(\ref{rem:prop}(\ref{item:2.7.1}))}
           \end{longtable}

\medskip\noindent
(c)~$({-})^*$ is the unique Elgot algebra structure
           such that $c$ is solution-preserving: in fact, for any 
           such Elgot algebra structure $({-})^*$
           and for any finitary flat equation
           morphism 
           $e:X\to HX+B$ we have
           \iffull $$\else $\fi
           c\cdot\Sol{e}=(c\after e)^*
           \iffull\,.$$\else $. \fi
           In particular, this is true for any equation morphism 
           of the form
           $$
           \xy
           \xymatrix@C+1pc{
           (s\after e')\equiv X
           \ar[r]^-{e'}
           &
           HX+C
           \ar[r]^-{HX+s}
           &
           HX+B
           }
           \endxy
           $$
           Thus, we conclude 
           $$
           \begin{array}{rcl@{\qquad}l}
           e^* 
           &=& ((c\cdot s)\after e)^* 
             & \mbox{($c\cdot s=\id$ and~\ref{rem:prop}\refeq{item:2.7.4})} \\
           &=& (c\after (s\after e))^* 
             & \mbox{(\ref{rem:prop}(\ref{item:2.7.2}))}\\
           &=& c\cdot \Sol{(s\after e)} 
             & \mbox{($c$ preserves solutions)}
           \end{array}
           $$

\medskip\noindent
(d)~$c$ is a coequalizer of $f$ and $g$ in $\Elgot{H}$. In fact,
           let $h:(B,\beta,\Sol{({-})})\to (D,\delta,({-})^+)$ be a 
           solution-preserving morphism with $h\cdot f=h\cdot g$. There is
           a unique homomorphism $\ol{h}:C\to D$ of $H$-algebras with
           $\ol{h}\cdot c=h$ (because $c$ is a coequalizer of $f$ and $g$
           in $\Alg H$). We prove that $\ol{h}$ is solution-preserving.
           Let $e:X\to HX+C$ be a finitary flat equation
           morphism. Then we have
           \iffull
           \setlongtables
           \begin{longtable}{>{$}r<{$}>{$}c<{$}>{$}l<{$}l}
             \endhead
             \endfoot
           \else
           $$
           \begin{array}{rclp{6cm}}
             \fi
           \ol{h}\cdot e^* 
           &=& \ol{h}\cdot c\cdot \Sol{(s\after e)}
             & \mbox{(Definition of $({-})^*$)} \\
           &=& h\cdot \Sol{(s\after e)} 
             & \mbox{($h=\ol{h}\cdot c$)} \\
           &=& (h\after (s\after e))^+
             & \mbox{($h$ preserves solutions)}\\
           &=& ((h\cdot s)\after e)^+ 
             & \mbox{(\ref{rem:prop}(\ref{item:2.7.2}))} \\
           &=& ((\ol{h}\cdot c\cdot s)\after e)^+ 
             & \mbox{($h=\ol{h}\cdot c$)} \\
           &=& (\ol{h}\after e)^+
             & \mbox{($c\cdot s=\id$)}     
           \iffull
           \end{longtable}
           \else
           \end{array}
           $$
           \fi
           \noindent
           as desired.
This completes the proof.
\end{proof}

\begin{exa}\label{ex:jsl2}
  Let $A$ be a join semilattice. Recall from Example~\ref{ex:jsl} the function
  $\alpha: RA \to A$ assigning to a rational binary tree $t$ in $RA$ the join
  of the labels of all leaves of $t$ in $A$. Since joins commute with joins
  it follows that this is the structure of an Eilenberg-Moore algebra on $A$.
  Thus, $A$ is an Elgot algebra as described in Example~\ref{ex:jsl}.
\end{exa}

\section{Complete Elgot Algebras}
\label{sec:5}
\setcounter{equation}{0}

\comment{
\begin{rem}
  \label{rem:5.1}
}
  Recall our standing assumptions that $H$ is an endofunctor of a category
  $\kat{A}$ with finite coproducts. Stefan Milius~\cite{m} has established
  that for every object-mapping $T$ of $\kat{A}$ the following three statements
  are equivalent:
  \begin{enumerate}
    \renewcommand{\labelenumi}{(\alph{enumi})}
  \item for every object $Y$, $TY$ is a final coalgebra for $H(-) + Y$,
  \item for every object $Y$, $TY$ is a free completely iterative $H$-algebra
    on $Y$, and
  \item $T$ is the object part of a free completely iterative monad $\T$ on
    $H$. 
  \end{enumerate}
  See also~\cite{aamv}, where the monad $\T$ is described and the implication
  that~(a) implies~(c) is proved.
  
  We are going to add another equivalent item to the above list,
  bringing complete Elgot algebras into the picture. The statements~(a) to~(c)
  are equivalent to
  \begin{enumerate}
    \renewcommand{\labelenumi}{(\alph{enumi})}
    \setcounter{enumi}{4}
  \item[(d)] for every  object $Y$, $TY$ is a free complete Elgot algebra on $Y$.
  \end{enumerate} 
  Furthermore, recall from~\cite{aamv} that $H$ is {\em iteratable} if there
  exist objects $TY$ such that one of the above equivalent statements holds. We
  will describe for every iteratable endofunctor the category $\kat{A}^\T$ of
  Eilenberg--Moore algebras---it is isomorphic to the category of complete
  Elgot algebras for $H$.  \comment{
\end{rem}
}

\begin{exa}
  For a polynomial endofunctor $H_\Sigma$ of $\Set$, the above monad
  $\T$ is the
  monad $T_\Sigma$ of all (finite and infinite) $\Sigma$-trees. 
\end{exa}

In the following result the concept of \emph{solution-preserving morphism} is
defined for complete Elgot algebras analogously to
Definition~\ref{dfn:solpres}: the equation~\refeq{eq:solpres} holds for
\emph{all} flat equation morphisms $e$. We denote by
$$\CElgot{H}$$
the category of all complete Elgot algebras and solution-preserving
morphisms. 

\begin{lem}\label{lem:compsol}
  Every solution-preserving morphism between complete Elgot algebras is a
  homomorphism of $H$-algebras.
\end{lem}
\begin{rem}
  If the base category $\A$ is locally finitely presentable and $H$ is
  finitary, then this lemma is a special case of Lemma~\ref{lem:solpres}.
  However, the statement of Lemma~\ref{lem:compsol} is more general, and the
  proof is completely different.
\end{rem}
\iffull
  \begin{proof}[Proof of Lemma~\ref{lem:compsol}]
    Let $(A, \alpha, \solop)$ and $(B, \beta, \Solop)$ be complete Elgot
    algebras. 
    Suppose that $h: A \to B$ is a solution-preserving morphism, and consider
    the flat equation morphism
    $$
    e \equiv
    \xymatrix@1@C+2pc{
      HA + A \ar[r]^-{H\inr + A} & H(HA + A) + A
      }\,.
    $$
    Its solution fulfils $\sol{e} = [\alpha, A]: HA + A \to A$. In fact, the
    following diagram
    $$
    \xymatrix@C+1pc{
      HA + A \ar[r]^-{\sol{e}} \ar[d]_{H\inr + A} & A \\
      H(HA + A) + A \ar[r]_-{H\sol{e} + A} & HA + A \ar[u]_{[\alpha, A]}
      }
    $$
    commutes. Thus, $\sol{e} \cdot \inr = \id$, and then it follows that
    $\sol{e}\cdot\inl = \alpha$. 
    Since $h$ preserves solutions we know that $h \cdot \alpha$ is the
    left-hand component of the solution of the following
    flat equation morphism
    $$
    h \after e \equiv 
    \xymatrix@1@C+3pc{
      HA + A \ar[r]^-{H\inr + A} & H(HA + A) + A \ar[r]^-{H(HA + A) + h} & 
      H(HA + A) + B\,;
      }
    $$
    in symbols, $\Sol{(h \after e)} \cdot \inl = h \cdot \alpha$.
    Now consider the diagram
    $$
    \xymatrix@+1pc{
      HA + A 
      \ar[r]^-{H\inr + A}
      \ar[d]_{Hh + h}
      \ar[rrrd]^{H(\inr \cdot h) + h}
      &
      H(HA + A) + A
      \ar[rr]^-{H(HA + A) + h}
      &
      &
      H(HA + A) + B
      \ar[d]^{H(Hh + h) + B}
      \ar@{<-} `u[l] `[lll]_{h \after e}
      \\
      HB + B
      \ar[rrr]_-{H\inr + B}
      & 
      &
      &
      H(HB + B) + B
      }
    $$
    which trivially commutes. Hence, $Hh + h$ is a morphism of equations from
    $h \after e$ to $H\inr + B$. 
    By a similar argument as for $\sol{e}$ above we obtain
    $[\beta, B] = \Sol{(H\inr + B)}$. Thus, by functoriality we conclude
    that $h$ is an $H$-algebra homomorphism:
    $$
    h\cdot \alpha 
    = 
    \Sol{(h \after e)} \cdot \inl 
    = 
    [\beta, B] \cdot (Hh + h) \cdot \inl
    = 
    \beta \cdot Hh\,.
    $$
    This completes the proof.
  \end{proof}
\fi
\comment{
\begin{rem}
  At the end of the current Section we will show that complete Elgot algebras
  are precisely the Eilenberg--Moore algebras of the completely iterative monad
  $\T$, i.\,e., $\kat{A}^\T$ and $\CElgot H$ are isomorphic categories. 
  
  In order to do that we will now relate final coalgebras (or equivalently,
  free CIAs) to free complete Elgot algebra: we prove the equivalence of~(d)
  to~(a) through~(c) in~\ref{rem:5.1}. We denote by $U: \CElgot{H} \to
  \kat{A}$ the natural forgetful functor. 
\end{rem}
 }
\begin{thm}\label{thm:elgot-cia}
  Let $Y$ be an object of $\kat{A}$. Then the following are equivalent:
  \begin{enumerate}
    \renewcommand{\labelenumi}{(\arabic{enumi})}
  \item $TY$ is a final coalgebra for $H(-) + Y$, and
  \item $TY$ is a free complete Elgot algebra on $Y$.
  \end{enumerate}
\end{thm}
\iffull
Before we prove this theorem, we need a technical lemma:
\begin{construction}
  Let $(A, \alpha, \solop)$ be a complete Elgot algebra. For every morphism $m:
  Y \to A$ we construct a new complete Elgot algebra on $HA + Y$ as follows:
  \begin{enumerate}
  \item The algebra structure is
    $$
    \xymatrix@1@C+1pc{
      H(HA + Y) \ar[r]^-{H[\alpha, m]} & HA \ar[r]^-\inl & HA + Y\,.
      }
    $$
  \item The choice $\Solop$ of solutions is as follows: for every flat equation
    morphism $e: X \to HX + HA + Y$ consider the flat equation morphism
    $$
    \ol e  \equiv
    \xymatrix@1{
      X \ar[r]^-e & HX + HA + Y \ar[rr]^-{HX + [\alpha, m]} & & HX + A\,,
      }
    $$
    and put
    $$
    \Sol{e} \equiv 
    \xymatrix@1{
      X \ar[r]^-e & HX + HA + Y \ar[rr]^-{[H\sol{\ol e}, HA] + Y} & & 
      HA + Y\,.
      }
    $$
    Notice that $\ol e = [\alpha, m] \after e$. 
  \end{enumerate}
\end{construction}
\begin{lem}\label{lem:constr}
  The above construction defines a complete Elgot algebra such that $[\alpha,
  m]: HA + Y \to A$ is a solution-preserving morphism into the original algebra.
\end{lem}
\begin{proof}
  (1)~The morphism $[\alpha, m]$ is solution-preserving: In fact, for any flat
  equation morphism $e: X \to HX + HA + Y$ we have the following commutative
  diagram
  $$
  \xymatrix@R-7pt{
    \rule{0pt}{10pt}
    X \ar[r]^-e \ar[rd]_{\ol e} \ar `u[r] `[rrr]^-{\Sol e} [rrr] 
    \ar `d[ddr] [ddrrr]_{\sol{\ol e} = \sol{([\alpha, m] \after e)}} &
    HX + HA + Y \ar[rr]^-{[H\sol{\ol e}, HA] + Y} \ar[d]^{HX + [\alpha, m]} & &
    HA + Y \ar[dd]^{[\alpha, m]} \\
    & HX + A \ar[r]_-{H\sol{\ol e} + A} & HA + A \ar[rd]_(.4){[\alpha, A]} \\
    & & & A
    }
  $$
  The lower left-hand part commutes since $\sol{\ol e}$ solves $\ol e$; the
  upper part is the definition of $\Solop$, the left-hand triangle is the
  definition of $\ol e$, and all components of the inner right-hand part are
  clear. 
  
  \medskip\noindent
  (2)~The morphism $\Sol{e}$ is a solution of $e$. In fact, the following
  diagram
  $$
  \let\objectstyle=\scriptstyle
  \let\labelstyle=\scriptscriptstyle
  \xymatrix@C+1.8pc{
    \rule{0mm}{10pt}X \ar[r]^-e \ar[d]_e \ar `u[r] `[rrr]^{\Sol{e}} [rrr] & 
    HX + HA + Y\ar[rr]^-{[H\sol{\ol e}, HA] + Y} \ar[d]_{He + HA + Y} & &
    HA + Y
    \\
    \rule[-4pt]{0pt}{1pt}
    HX + HA + Y \ar[r]_-{He + HA + Y} \ar `d[r] `[rrr]_{H \Sol{e} + HA + Y} [rrr] &
    H(HX + HA + Y) + HA + Y \ar[rr]_-{H ([H\sol{\ol e}, HA] + Y) + HA + Y} & &
    H(HA + Y) + HA + Y\ar[u]^{[\inl \cdot H[\alpha, m], HA + Y]} 
    }
  $$
  commutes: the upper and lower part as well as the left-hand square are
  obvious, and so are the middle and right-hand components of the right-hand
  square. To see that the left-hand component commutes, we remove $H$ and
  observe that the following diagram commutes:
  $$
  \xymatrix{
    X \ar[rrr]^-{\sol{\ol e}} \ar[dd]_e \ar[rd]^-{\ol e} & & &     
    A 
    \\
    & HX + A \ar[r]_-{H\sol{\ol e} + A} & 
    HA + A \ar[ru]^-{[\alpha, A]} 
    \\
    HX + HA + Y \ar[ru]_(.7)*+{\labelstyle HX + [\alpha, m]} \ar[rrr]_-{[H\sol{\ol e}, HA] + Y}
    & & & 
    HA + Y \ar[uu]_{[\alpha, m]}
    }
  $$

  \medskip\noindent
  (3)~Functoriality: Suppose we have a morphism $h: e\to f$ of equations. Then
  $h: \ol{e} \to \ol{f}$ is also one, and we obtain the following diagram
  $$
  \xymatrix@C+2pc{ 
    X \ar[r]^-e \ar[dd]_h \ar `u[r] `[rr]^{\Sol{e}} [rrd] &
    HX + HA + Y \ar[dd]_{Hh + HA + Y} 
    \ar[rd]^*+{\scriptstyle [H\sol{\ol e}, HA] + Y} &
    \\
    & & HA + Y
    \\
    Z \ar[r]_-f \ar `d[r] `[rr]_{\Sol{f}} [rru] & HZ + HA + Y
    \ar[ru]_*+{\scriptstyle [H\sol{\ol f}, HA] + Y} & 
    }
  $$
  It commutes: in the triangle the components with domains $HA$ and $Y$ are
  clear, for the left-hand component remove $H$ and use functoriality of
  $\solop$, and all other parts are obvious. 

  \medskip\noindent
  (4)~Compositionality: Suppose we have two flat equation morphisms
  $$f: X \to HX + Z \quad \textrm{and} \quad g: Z \to HZ + HA + Y\,.$$
  Observe that $\Sol{(\Sol g \after f)}$ is the following morphism
  \begin{equation}
    \label{eq:cia1}
    \vcenter{
        \let\objectstyle=\scriptstyle
        \let\labelstyle=\scriptscriptstyle
    \xymatrix@C+1pc{
      X 
      \ar[r]^-f 
      & 
      HX + Z 
      \ar[r]^-{HX + g} 
      & 
      HX + HZ + HA + Y 
      \ar[rr]^-{HX + [H\sol{\ol g}, HA] + Y} 
      \ar[rrd]_{\left[H\left(\sol{\ol{\Sol g \after f}}\right), H\sol{\ol g},
          HA\right] + Y \hspace*{35pt}}   
      & & 
      HX + HA 
      \ar[d]^{\left[H\left(\sol{\ol{\Sol g \after f}}\right), HA\right] + Y} 
      \ar@{<-} `u[l] `[llll]_{\Sol g \after f} [llll]
      \\
      & & & & HA
      }}
  \end{equation}
  and $\Sol{(g \plus f)} \cdot \inl$ is the following morphism
  \begin{equation}
    \label{eq:cia2}
    \vcenter{
        \let\objectstyle=\scriptstyle
        \let\labelstyle=\scriptscriptstyle
    \xymatrix@C+1pc{
      X 
      \ar[r]^-f 
      & 
      HX + Z 
      \ar[r]^-{HX + g} 
      & 
      HX + HZ + HA + Y 
      \ar[rr]^-{\can + HA + Y} 
      \ar[rrd]_*+{\scriptstyle \left[H\left(\sol{(\sol{\ol g} \after f)}\right),
          H\sol{\ol g}, HA\right] + Y \hspace*{25pt}} 
      & & 
      H(X+Z) + HA + Y 
      \ar[d]^{\left[H\left(\sol{\ol{g \smallplus f}}\right), HA\right] + Y} 
      \ar@{<-} `u[l] `[llll]_{g \smallplus f} [llll]
      \\
      & & & & HA + Y
      }}
  \end{equation}
  In fact, to see that the last triangle commutes consider the components
  separately. The right-hand one with domain $HA + Y$ is trivial, and for the
  left-hand one with domain $HX + HZ$ it suffices to observe the following
  equations:
  $$
  \begin{array}{rcl@{\qquad}l}
    \sol{\ol{g \plus f}} & = & \sol{([\alpha, m] \after (g \plus f))} &
    \mbox{(Definition of $\ol{g \plus f}$)} \\
    & = & \sol{(([\alpha, m] \after g) \plus f)} &
    \mbox{(\ref{rem:prop}\refeq{item:2.7.4})} \\
    & = & \left[(\sol{\sol{([\alpha, m] \after g)} \after f)}, 
      \sol{([\alpha, m] \after g)}\right] &
    \mbox{(by~\refeq{eq:modCIA})} \\
    & = & \left[\sol{(\sol{\ol g} \after f)}, \sol{\ol g}\right] &
    \mbox{(Definition of $\ol{g}$)} \rule{0mm}{15pt}
  \end{array}
  $$
  To show the desired identity of the morphisms in~\refeq{eq:cia1}
  and~\refeq{eq:cia2} it suffices to prove that the slanting arrows in
  those diagrams are equal. The last three components are clear, and for the
  first one the following equations are sufficient:
  $$
  \begin{array}{rcl@{\qquad}l}
    \sol{\ol g} \after f & = & \sol{([\alpha, m] \after g)} \after f & 
    \mbox{(Definition of $\ol{g}$)} \\
    & = & ([\alpha, m] \cdot \Sol{g}) \after f & 
    \mbox{($[\alpha,m]$ preserves solutions)} \\
    & = & [\alpha, m] \after (\Sol{g} \after f) &
    \mbox{(\ref{rem:prop}\refeq{item:2.7.2})} \\
    & = & \ol{\Sol g \after f} & 
    \mbox{(Definition of $\ol{\Sol g \after f}$)}
  \end{array}
  $$
  This completes the proof. 
\end{proof}

\begin{proof}(Theorem~\ref{thm:elgot-cia}.)
  By Theorems~2.8 and~2.10 of~\cite{m},
  statement~(1) is equivalent to
  
  \smallskip\noindent
  (1') $TY$ is a free CIA on $Y$,
  
  \smallskip\noindent
  We prove now that~(2) is equivalent to~(1)
  by showing the implications $(1')\Rightarrow (2)\Rightarrow (1)$.
  We first observe that for a free complete Elgot
  algebra on $Y$, $(TY, \tau_Y, \solop)$, with a universal arrow $\eta_Y: Y \to
  TY$, the morphism $[\tau_Y, \eta_Y]: HTY + Y \to TY$ is an isomorphism.  In
  fact, by Lemma~\ref{lem:constr}, $HTY + Y$ carries the
  complete Elgot algebra structure
  and $j = [\tau_Y, \eta_Y]$ is solution-preserving and
  fulfils $j \cdot \inr = \eta_Y$. Invoke the freeness of $TY$ to obtain a unique
  solution-preserving morphism $i: TY \to HTY + Y$ such that $i \cdot \eta_Y =
  \inr$. It follows that $j \cdot i = \id$. By Lemma~\ref{lem:compsol}, $i$ is an
  $H$-algebra homomorphism. Thus the following square
  $$
  \xymatrix{
    HTY + Y \ar[r]^-j \ar[d]_{H i + Y} & TY \ar[d]^i \\
    H(HTY + Y) + Y \ar[r]_-{Hj + Y} & HTY + Y 
    }
  $$
  commutes, whence $i \cdot j = \id$. 
  
  \medskip\noindent 
  Proof of (2) $\Rightarrow$ (1). Let $(TY, \tau_Y, \solop)$
  be a free complete Elgot algebra on $Y$ with a universal arrow $\eta_Y: Y
  \to TY$. Then $[\tau_Y, \eta_Y]: HTY + Y \to TY$ is an isomorphism with an
  inverse $i$. We prove that $(TY, i)$ is a final coalgebra for $H(-) + Y$. So
  let $c: X \to HX + Y$ be any coalgebra, and form the flat equation morphism
  \begin{equation}\label{eq:e}
    e \equiv
    \xymatrix@1{
      X \ar[r]^-c & HX + Y \ar[rr]^-{HX + \eta_Y} & & HX + TY\,.
      }
  \end{equation}
  Then $\sol{e}$ is a coalgebra homomorphism from $(X, c)$ to $(TY, i)$; in
  fact, it suffices to establish that the diagram
  $$
  \xymatrix@C+1pc{
    X 
    \ar[rr]^-c
    \ar[rd]^e
    \ar[ddd]_{\sol{e}}
    &
    &
    HX + Y
    \ar[ddd]^{H\sol e + Y}
    \ar[ld]_(.6)*+{\labelstyle HX + \eta_Y}
    \\
    &
    HX + TY
    \ar[d]^{H\sol e + TY}
    \\
    &
    HTY + TY
    \ar[ld]_*+{\labelstyle [\tau_Y, TY]}
    \\
    TY 
    &
    &
    HTY + Y
    \ar[ll]^{[\tau_Y, \eta_Y] = i^{-1}}
    \ar[lu]_*+{\labelstyle HTY + \eta_Y}
    }
  $$
  commutes. The upper part is~\refeq{eq:e}, the left-hand part commutes
  since $\sol{e}$ is a solution of $e$, the right-hand one commutes
  trivially,
  and the lower part is obvious.

  Now suppose that $s$ is a coalgebra homomorphism from $(X,c)$ to $(TY, i)$.
  We prove that $s = \sol{e}$. Observe first that $s$ is a
  morphism of equations from $e$ to the following flat equation morphism
  \begin{equation}
    \label{eq:f}
    f \equiv
    \xymatrix@1{
      TY \ar[r]^-i & HTY + Y  \ar[rr]^-{HTY + \eta_Y} & & HTY + TY\,,
      }
  \end{equation}
  In fact, the following diagram 
  $$
  \xymatrix{
    X 
    \ar[r]^-c
    \ar[d]_s
    &
    HX + Y
    \ar[rr]^-{HX + \eta_Y}
    \ar[d]^{Hs + Y}
    & & 
    HX + TY
    \ar[d]^{Hs + TY}
    \ar@{<-} `u[l] `[lll]_-e [lll]  
    \\
    TY 
    \ar[r]_-i 
    &
    HTY + Y 
    \ar[rr]_-{HTY + \eta_Y} 
    & 
    &
    HTY + TY 
    \ar@{<-} `d[l] `[lll]^f [lll] 
    }
  $$
  commutes: the left-hand square does since $s$ is a coalgebra homomorphism, 
  the right-hand one commutes trivially and the upper and lower parts
  are due to~\refeq{eq:e} and~\refeq{eq:f}. 
  By functoriality of $\solop$ we obtain $\sol f \cdot s =
  \sol{e}$. We shall show below that $\sol{f}: TY \to TY$ is a
  solution-preserving map with $\sol{f} \cdot \eta_Y = \eta_Y$. By the freeness of
  $TY$, we then conclude that $\sol{f} = \id$, whence $\sol{e} = s$ as desired.

  To see that $\sol{f} \cdot \eta_Y = \eta_Y$ consider the following 
  diagram
  $$
  \xymatrix@C+1pc{
    \rule{0pt}{10pt}
    TY \ar[d]_{\sol f} \ar@<.5ex>[r]^-i \ar `u[r] `[rr]^f [rr] & 
    HTY + Y \ar@<.5ex>[l]^-{[\tau_Y,\eta_Y]} \ar[r]^-{HTY + \eta_Y} & 
    HTY + TY \ar[d]^{H \sol f + TY} \\
    TY & & HTY + TY \ar[ll]^-{[\tau_Y, TY]}
    }
  $$
  which commutes since $\sol f$ is a solution of $f$. Follow the right-hand
  component of the coproduct $HTY + Y$ to see the desired equation.

  To complete our proof we must show that the following triangle
  \begin{equation}
    \label{eq:solf}
    \vcenter{
      \xymatrix{
        & X \ar[ld]_{\sol{e}} \ar[rd]^{\sol{(\sol f \after e)}} \\
        TY \ar[rr]_{\sol{f}} & & TY
        }}    
  \end{equation}
  commutes for any flat equation morphism $e: X \to HX + TY$.
  Notice first that
  \begin{equation}\label{eq:cia}
    \sol{(\sol f \after e)} = \sol{(f \plus e)} \cdot \inl: X \to TY
  \end{equation}
  by compositionality. Furthermore, we have
  a morphism $[\sol{e},TY]$ of equations from
  $f \plus e$ to $f$. In fact, the diagram below commutes:
  $$
      \xymatrix@C+1pc{
        X+TY \ar[dd]|*+{\labelstyle [\sol{e}, TY]} \ar[r]^-{[e, \inr]} 
        & 
        HX + TY \ar[d]_{H \sol{e} + TY} \ar[r]^-{HX + i} & 
        HX + HTY + Y \ar[r]^-{\can + \eta_Y} \ar[dd]|*+{\labelstyle [H\sol{e},
          HTY] + Y} &  
        H(X + TY) + TY \ar[dd]|*+{\labelstyle H[\sol{e}, TY] + TY} 
        \ar@{<-} `u[l] `[lll]_-{f \smallplus e} [lll] \\
        & HTY + TY \ar[ld]_{[\tau_Y, TY]} \ar[rd]^{[\inr, i]} \\
        TY \ar[rr]_-i & & 
        HTY + Y \ar[r]_-{HTY + \eta_Y} & 
        HTY + TY \ar@{<-} `d[l] `[lll]^-f [lll]
        }
  $$
  By functoriality we obtain the following equality
  $$
  \sol{f} \cdot [\sol{e}, TY] = \sol{(f \plus e)}\,,$$
  whose left-hand
  component proves due to~\refeq{eq:cia} the desired commutativity
  of~\refeq{eq:solf}.

  \sloppypar
  \medskip\noindent
  (1') $\Rightarrow$ (2). We only need to show the universal property. Suppose
  that 
  $(TY, \tau_Y, \solop)$ is a free CIA on $Y$ with a universal arrow $\eta_Y:
  Y \to TY$. Due to the equivalence of~(1) and~(1'), $[\tau_Y, \eta_Y]$ has an
  inverse $i$, and $(TY, i)$ is a final coalgebra for the functor $H(-) + Y$.
  Now let $(A, \alpha, \Solop)$ be a complete Elgot algebra and let $m: Y \to
  A$ be a morphism of $\kat{A}$. Solve the following flat equation morphism
  $$
  g \equiv
  \xymatrix{
    TY \ar[r]^-i & HTY + Y \ar[rr]^-{HTY + m} & & HTY + A
    }
  $$
  in $A$ to obtain a morphism $h = \Sol{g}: TY \to A$. We first check that
  $h \cdot \eta_Y = m$. In fact, the following diagram
  $$
  \xymatrix@C+1pc{
    \rule{0pt}{10pt}
    TY \ar[d]_h \ar@<.5ex>[r]^-i \ar `u[r] `[rr]^g [rr] & 
    HTY + Y \ar@<.5ex>[l]^-{[\tau_Y, \eta_Y]} \ar[r]^-{HTY + m} & 
    HTY + A \ar[d]^{Hh + A} \\
    A & & HA + A \ar[ll]^-{[\alpha, A]}
    }
  $$
  commutes since $h$ is a solution of $g$. Consider the right-hand
  component of the coproduct $HTY + Y$ to obtain the desired equation. 
  
  Next let us show that $h$ is a solution-preserving morphism.
  More precisely, we show
  that for any equation morphism $e: X \to HX + TY$ the triangle
  \begin{equation}\label{diag:trih}
    \vcenter{
      \xymatrix{
        & 
        X 
        \ar[ld]_{\sol{e}}
        \ar[rd]^{\Sol{(h \after e)}}
        \\
        TY
        \ar[rr]_-h
        &
        &
        A
        }
      }
  \end{equation}
  commutes. Since $h = \Sol{g}$, the equality
  \begin{equation}\label{eq:ciatwo}
    \Sol{(h\after e)} = \Sol{(g \plus e)} \cdot \inl: X \to A
  \end{equation}
  holds due to compositionality of $\Solop$. Moreover, $[\sol{e}, TY]$ is a
  morphism of equations from $g \plus e$ to $g$. In fact, consider the following
  commutative diagram
  $$
  \xymatrix@C+1pc{
    X+TY \ar[dd]_{[\sol{e}, TY]} \ar[r]^-{[e, \inr]} 
    &
    HX + TY \ar[d]_{H \sol{e} + TY} \ar[r]^-{HX + i} 
    & 
    HX + HTY + Y \ar[r]^-{\can + m} \ar[dd]^{[H\sol{e}, HTY] + Y} 
    & 
    H(X + TY) + A \ar[dd]^{H[\sol{e}, TY] + A} 
    \ar@{<-} `u[l] `[lll]_-{g \smallplus e} [lll] 
    \\
    & 
    HTY + TY \ar[ld]_{[\tau_Y, TY]} \ar[rd]^{[\inr, i]} 
    \\
    TY \ar[rr]_-i 
    & & 
    HTY + Y \ar[r]_-{HTY + m} 
    & 
    HTY + A \ar@{<-} `d[l] `[lll]^-g [lll]
    }
  $$
  By functoriality of
  $\Solop$ we obtain the equation
  $$
  \Sol{g} \cdot [\sol{e}, TY] = \Sol{(g \plus e)}
  $$
  whose left-hand component is the desired~\refeq{diag:trih}
  due to~\refeq{eq:ciatwo}. Thus, $h$ is solution-preserving.
  
  To show uniqueness suppose that $h: TY \to A$ is any solution-preserving
  morphism with $h \cdot \eta_Y = m$. Observe that we have $g = h \after f$,
  where $f$ is the flat equation morphism of~\refeq{eq:f}. Since $h$ preserves
  solutions we have
  $$\Sol{g} = \Sol{(h \after f)} = h \cdot \sol{f}\,.$$
  To complete the proof it
  suffices to show that $\sol{f} = \id$. This can be done with precisely the
  same argument as in the part $(2)\Rightarrow (1)$ of the present proof.
  One shows that $\sol f: TY \to TY$ is a
  solution-preserving morphism such that $\sol{f} \cdot \eta_Y = \eta_Y$. From the
  universal property of the free CIA $TY$ it follows that $\sol{f} = \id$, see
  also Proposition~2.3 in~\cite{m}.
\end{proof}
\fi

\iffull
\begin{cor}
  \label{cor:sumup}
  For any endofunctor $H:\kat{A} \to \kat{A}$
  the following are equivalent:
  \begin{enumerate}
    \renewcommand{\labelenumi}{(\arabic{enumi})}
  \item $H$ is iteratable, i.\,e., there exist final coalgebras for all
    functors $H(-) + Y$;
  \item there exist free completely iterative $H$-algebras on every object
    $Y$;
  \item there exist free complete Elgot algebras on every object $Y$. 
  \end{enumerate}
\end{cor}
\begin{proof}
  See~\cite{m}, Corollary~2.11 for $(1) \Leftrightarrow (2)$. The equivalence
  $(2) \Leftrightarrow (3)$ follows from Theorem~\ref{thm:elgot-cia}. 
\end{proof}
\fi

\comment{
\begin{cor}
  The following are equivalent:
  \begin{enumerate}
  \item $T$ is a final $H$-coalgebra,
  \item $T$ is an initial CIA, and
  \item $T$ is an initial complete Elgot algebra.
  \end{enumerate}
\end{cor}
}
\comment{ 
  What is meant is that $\alpha: T \to HT$ is a final coalgebra if and only
  if $\alpha^{-1}: HT \to T$ is an initial CIA, and equivalently, an initial
  complete Elgot algebra.  The proof follows from Theorem~\ref{thm:elgot-cia}
  applied to the initial object $Y = 0$ of $\kat{A}$.
}

\begin{thm}
  \label{thm:c-emalg}
  If $H$ is an iteratable functor, then the category $\CElgot{H}$ of complete
  Elgot algebras is isomorphic to the Eilenberg--Moore category $\kat{A}^\T$ of
  $\T$-algebras for the free completely iterative monad $\T$ of $H$.
\end{thm}
\begin{proof}
  \iffull
  By Corollary~\ref{cor:sumup}, 
  \else
  By Theorem~\ref{thm:elgot-cia}, 
  \fi
  the natural forgetful functor $U: \CElgot{H}
  \to \kat{A}$ has a left adjoint $Y \mapsto TY$. Thus, the monad obtained by
  this adjunction is $\T$. To prove that the comparison functor $K: \CElgot{H}
  \to \kat{A}^\T$ is an isomorphism use Beck's Theorem. In fact, the argument
  that $U$ creates coequalizers of $U$-split pairs is entirely analogous to
  that of Theorem~\ref{th:4.6}.
\end{proof}

\begin{exa}\label{ex:prooflattice}
  Let $A$ be a complete lattice. Recall from Example~\ref{ex:lat} the function
  $\alpha: TA \to A$ assigning to every binary tree $t$ in $TA$ the join of all
  labels of leaves of $t$ in $A$. Since joins commute with joins it follows
  that $\alpha: TA \to A$ is the structure of an Eilenberg-Moore algebra on
  $A$. Thus, $A$ is a complete Elgot algebra as described in
  Example~\ref{ex:lat}. 
\end{exa}

\comment{
As an application let us consider algebras of a locally continuous functor on
the category $\CPO_\bot$ of cpos with a least element and strict continous
maps. 

\begin{cor}
  Let $H:\CPO_\bot\to\CPO_\bot$ be a locally continuous functor. Every
  $H$-algebra $\alpha:HA\to A$ is a complete Elgot 
  algebra.
\end{cor}
\begin{proof}
  It is well-known, see~\cite{sp}, that since $H(-) + Y$ is locally continuous,
  a terminal coalgebra, $TY$, coincides with the initial algebra---i.\,e.,
  $TY$ is a free $H$-algebra on $Y$. Consequently, $\T$ is the free algebra
  monad of $H$ whose Eilenberg-Moore category is isomorphic to $\Alg H$. The
  desired result follows now from Theorem~\ref{thm:c-emalg}.
\end{proof}
}

\section{Summary and Future Work}
\label{sec:6}
\setcounter{equation}{0}

In this paper we introduce Elgot algebras: these are algebras
in which finitary flat equation morphisms have solutions satisfying two simple
axioms, one for change of parameters and one for simultaneous recursion.
Analogously, complete Elgot algebras are algebras in which flat equation
morphisms (not necessarily finitary) have solutions subject to the same two
axioms. These axioms are strikingly simple and have a clear intuitive
meaning.

Moreover, the motivation for Elgot algebras is provided canonically
by Elgot's iterative theories: Elgot algebras
are precisely the Eilenberg--Moore algebras for the free iterative theory
(as described by Calvin Elgot et al.~for signatures in~\cite{ebt} and by
the present authors~\cite{amv1,amv} for general endofunctors).
Analogously, complete Elgot
algebras are precisely the Eilenberg--Moore algebras for the free
completely iterative monad of
Calvin Elgot et al.~\cite{ebt} (generalized by
Peter Aczel and the present authors~\cite{aamv}, see also the work of
Stefan Milius~\cite{m}).

The assignment $e\mapsto \sol{e}$, which forms an Elgot algebra
structure, extends canonically from the above flat equation morphisms
$e$ to a much broader class of ``rational'' equation
morphisms. In that sense one gets close to iteration algebras of
Zolt\'{a}n \'{E}sik~\cite{esik}.
The relationship of the latter to Elgot algebras needs
further investigation.

One reason for presenting Elgot algebras not only in $\Set$ but
in general locally finitely presentable categories
is the fact that for the important
class of algebraic trees of Bruno
Courcelle~\cite{c}
(i.\,e., for the trees obtained by tree unfoldings of
recursive program schemes)
no abstract treatment has been presented so far.
We believe that algebraic trees can be treated abstractly
when working in the category $\Fin{\Set}$,
the locally finitely presentable category
of all finitary endofunctors of $\Set$.

Finally, our paper can be considered as part of a program proposed
by Larry Moss to rework the theory of recursive program schemes
and their semantics using coalgebraic methods.
Stefan Milius and Larry Moss~\cite{mm} introduce a general notion of recursive
program scheme and prove that any guarded recursive program scheme
has a unique ``uninterpreted'' solution in the final coalgebra for the functor
describing the given operations. For interpreted
semantics of recursive program schemes one needs a ``suitable'' notion
of an algebra. It is proved in ~\cite{mm} that
for every recursive program scheme an interpreted
solution can be given in any complete Elgot
algebra. As an application one obtains the classical theory of recursive
program schemes interpreted in
continuous or completely metrizable algebras. New
applications include, for example, recursively defined operations satisfying
extra conditions like commutativity, or applications
pertaining to non-well founded sets
or fractals.

\section*{Acknowledgement}
The authors would like to thank the anonymous
referees for their valuable comments.
                        
%
%


\comment{
%
%
\clearpage
\begin{appendix}
\section{Proofs}

\comment{ 
\subsection{Proof for Example~\ref{ex:3.2}} 

\begin{proof}
      The functoriality follows from the fact
      that given a homomorphism
      $$
      \xy
      \xymatrix{
      X
      \ar[0,2]^-{e}
      \ar[1,0]_{h}
      &
      &
      X+A
      \ar[1,0]^{h+A}
      \\
      X'
      \ar[0,2]_-{e'}
      &
      &
      X'+A
      }
      \endxy
      $$
      and a variable $x_i\in X$, then there are variables
      $p_0,\dots,p_k\in X$ as above if and only if there are
      variables $q_0,\dots,q_k\in X'$ with $e'(q_0)=q_1$, \dots,
      $e'(q_{k-1})=q_k$ and $e'(q_k)\in A$.

      To verify compositionality, let $e:X\to X+Y$
      and $f:Y\to Y+A$ be given. Consider the left-hand
      component of the solution $\sol{(f\plus e)}$ and suppose
      that $\sol{(f\plus e)}(x_j)=\alpha^k a$ for some variable
      $x_j\in X$. 
      The following are easily seen to be equivalent:
      \begin{enumerate}
      \item There are variables $x_j=p_0$, $p_1$, \dots, $p_k$ in $X+Y$ with
        $(f\plus e) (p_0)=p_1$, \dots, $(f\plus e)(p_{k-1})=p_k$ and $(f\plus
        e)(p_k)=a\in A$.
      \item There exists some $l\leq k$ such that $p_0,\dots,p_{l-1}\in X$,
        $p_l,\dots,p_k\in Y$ with $e(p_0)=p_1$, \dots, $e(p_{l-2})=p_{l-1}$ and
        $e(p_{l-1})=p_l\in Y$, and with $f(p_l)=p_{l+1}$, \dots,
        $f(p_{k-1})=p_k$, and $f(p_k)=a\in A$.
      \end{enumerate}
      The latter equations say precisely that $\sol{f}(p_l)=\alpha^{k-l}a$.
      So equivalently, we have $(\sol{f}\after e)(p_0)=p_1$, \dots,
      $(\sol{f}\after e)(p_{l-2})=p_{l-1}$ and 
      $(\sol{f}\after e)(p_{l-1})=\alpha^{k-l}a$, whence
      $\sol{(\sol{f}\after e)}(x_j)=\alpha^k a$.

      The case where $\sol{(f\plus e)}(x_j)=a_0$ is clear.
      
      Therefore, $\sol{(f\plus e)}$ behaves on the variables from $X$ precisely
      as $\sol{(\sol{f}\after e)}$ does.
\end{proof}
}

\end{appendix}
}

\end{document}